\documentclass[10pt]{article}
\usepackage{times, amsmath}
\usepackage{graphicx}
\usepackage{algorithm}
\usepackage{algorithmic}
\usepackage{multirow}
\usepackage{epsfig}
\usepackage{cite}
\usepackage{color}
\usepackage[table]{xcolor}
\usepackage[scriptsize]{subfigure}

\usepackage[labelfont=bf,labelsep=period,justification=raggedright]{caption}

\makeatletter
\renewcommand{\@biblabel}[1]{\quad#1.}
\makeatother

\newcommand{\grey}{\cellcolor{black!80}}
\newcommand{\blue}{\cellcolor{blue!15}}
\newcommand{\red}{\cellcolor{red!15}}

\begin{document}

\begin{flushleft}
{\Large
\textbf{A Latent Parameter Node-Centric Model for Spatial Networks}
}
\\
Nicholas D. Larusso$^{\ast}$,
Brian E. Ruttenberg, 
Ambuj Singh
\\
Dept. of Computer Science, University of California, Santa Barbara
\\
$\ast$ E-mail: Corresponding nlarusso@cs.ucsb.edu
\end{flushleft}


\section*{Abstract}
\label{sec:abstract}

Spatial networks, in which nodes and edges are embedded in space, play a vital role in the study of complex systems. For example, many social networks attach geo-location information to each user, allowing the study of not only topological interactions between users, but spatial interactions as well. The defining property of spatial networks is that edge distances are associated with a cost, which may subtly influence the topology of the network. However, the cost function over distance is rarely known, thus developing a model of connections in spatial networks is a difficult task.

In this paper, we introduce a novel model for capturing the interaction between spatial effects and network structure. Our approach represents a unique combination of ideas from latent variable statistical models and spatial network modeling. In contrast to previous work, we view the ability to form long/short-distance connections to be dependent on the \textit{individual} nodes involved. For example, a node's specific surroundings (e.g. network structure and node density) may make it more likely to form a long distance link than other nodes with the same degree. To capture this information, we attach a latent variable to each node which represents a node's \textit{spatial reach}. These variables are inferred from the network structure using a Markov Chain Monte Carlo algorithm.
We experimentally evaluate our proposed model on $4$ different types of real-world spatial networks (e.g. transportation, biological, infrastructure, and social). We apply our model to the task of link prediction and achieve up to a $35\%$ improvement over previous approaches in terms of the area under the ROC curve. Additionally, we show that our model is particularly helpful for predicting links between nodes with low degrees. In these cases, we see much larger improvements over previous models.

\section{Introduction}
\label{sec:intro}


Network analysis has been successfully applied to several scientific fields of study including sociology~\cite{Airoldi2009, Bakshy2012, Wong2006}, information science~\cite{Albert2000, Barabasi1999}, and ecology~\cite{Fortuna2006, Olesen2007}. In many cases, the spatial configuration of nodes is paramount in analyzing a network as it plays a significant role in the formation and maintenance of links. Despite the important relationship between space and structure, many models and analyses are limited to only the network topology. Obviously such models fail to capture important \textit{spatial} properties inherent in the data~\cite{Daraganova2011, Ferretti2011, Hayashi2006}. For example, in transportation networks, it is more economical to create short links between nodes~\cite{Gastner2006, Guimera2004}. Similarly, users in a social network are more likely to form links based on physically proximity because they have more interaction opportunities~\cite{Metcalf2005, Wong2006}.

Although a plethora of spatial network models have been introduced in the literature (e.g.~\cite{Barthelemy2011, Cerina2012, Daraganova2012, Expert2011, Wong2006, Yook2002}), they implicitly assume that the link-cost function is a function only of distance. For instance, the exponential distance model~\cite{Cerina2012, Yook2002} defines the probability of node $i$ connecting to node $j$ as $p(A_{ij} = 1) = \frac{k_i k_j}{Z} exp(-d_{ij} / \hat{d})$, where the single parameter, $\hat{d}$, is set to the average pairwise distance between all nodes that share a link. Such models assume that the only node-specific influence on forming connections is the degree.
\begin{table}[th]
    \renewcommand{\arraystretch}{1.3}
    \centering
    {\small
    \begin{tabular}{ | l | l | l | l | l | l |}
        \hline
        \textbf{Name} & \textbf{Type} & \textbf{Nodes} & \textbf{Edges} & \textbf{Area} & \textbf{Index of dispersion} \\ \hline
	\textit{C. elegans} & Biological & $277$ & $1,918$ & $0.012 \mu m^2$ & 7.163 \\ \hline
	Gowalla & Social & $600$ & $340$ & $776,000$ km$^2$ & 23.098 \\ \hline
	Internet & Infrastructure & $501$ & $2,661$ & $809,000$ km$^2$ & 11.317 \\ \hline
	US Airline & Transportation & $476$ & $2,773$ & $16,140,695$ km$^2$ & 1.564 \\ \hline
    \end{tabular}
  }
  \caption{Properties of the real-world spatial network datasets we examine in this paper. The last column refers to the \textit{index of dispersion}, a measure of complete spatial randomness (CSR) of the nodes~\cite{Diggle2003}. Values close to $1$ indicate that the nodes are likely to be distributed uniformly over the space, whereas values greater than $1$ result from too little dispersion (e.g. nodes tend to cluster in space).}
  \label{tbl:datasets}
\end{table}

\begin{figure*}[ht]
 	\centering
 	\subfigure[\textit{C. elegans}]{
 	    \label{fig:dist_celegans}
 	    \centering
 	    \includegraphics[keepaspectratio, width=0.23\textwidth]{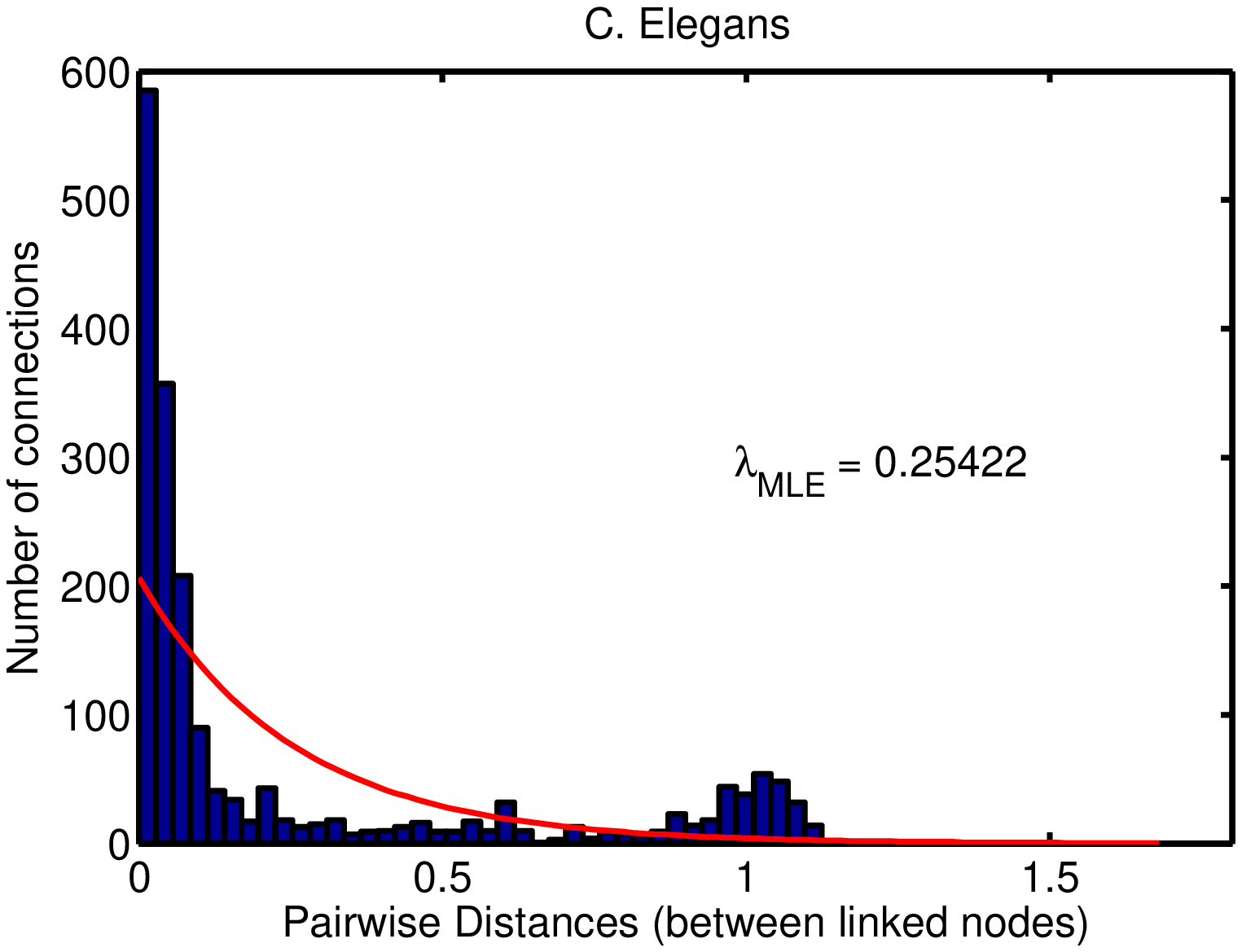}
 	}
 	\subfigure[Gowalla]{
 	    \label{fig:dist_gowalla}
 	    \centering
 	    \includegraphics[keepaspectratio, width=0.23\textwidth]{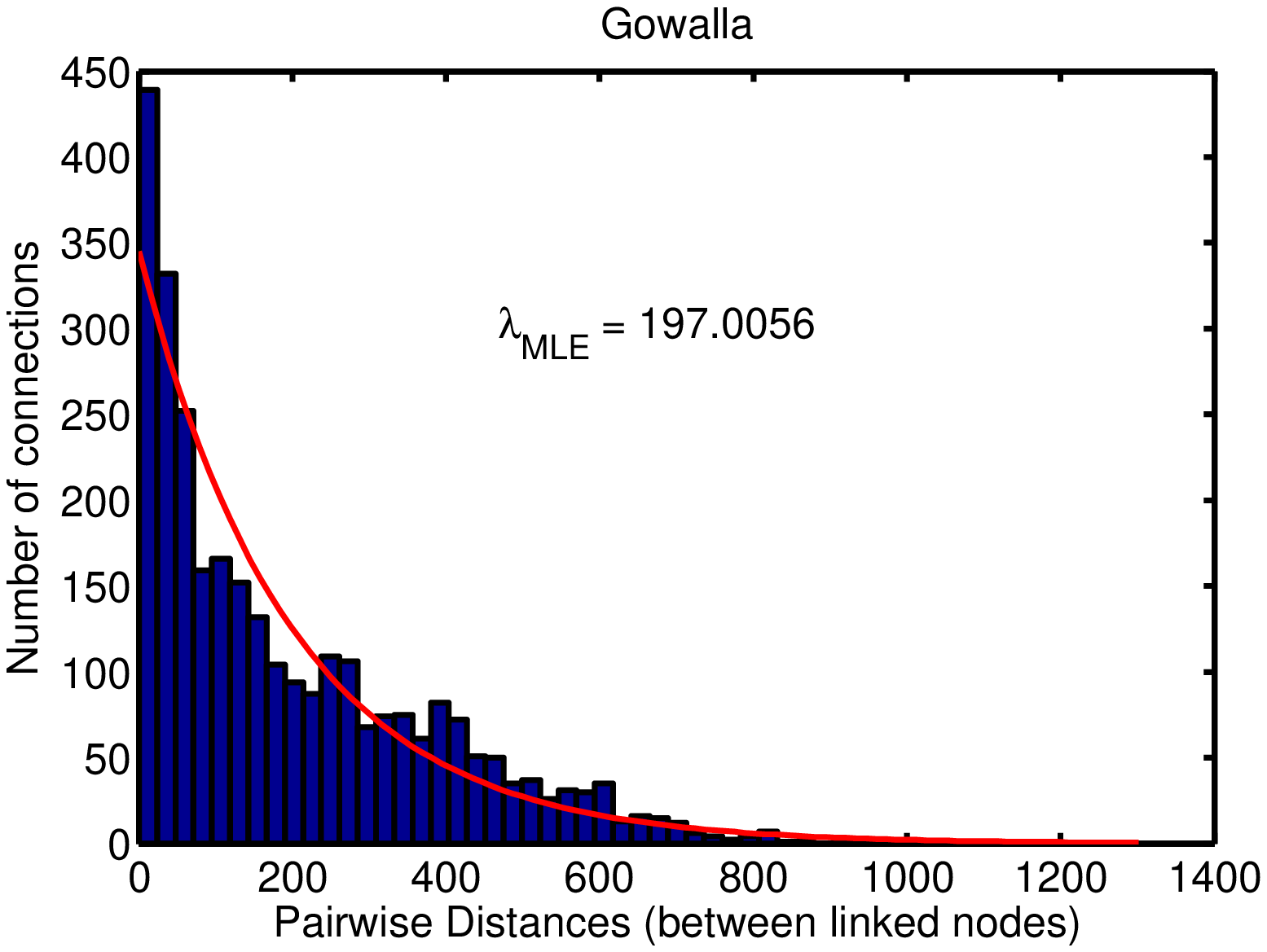}
 	}
 	\subfigure[CA Internet]{
 	    \label{fig:dist_internet}
 	    \centering
 	    \includegraphics[keepaspectratio, width=0.23\textwidth]{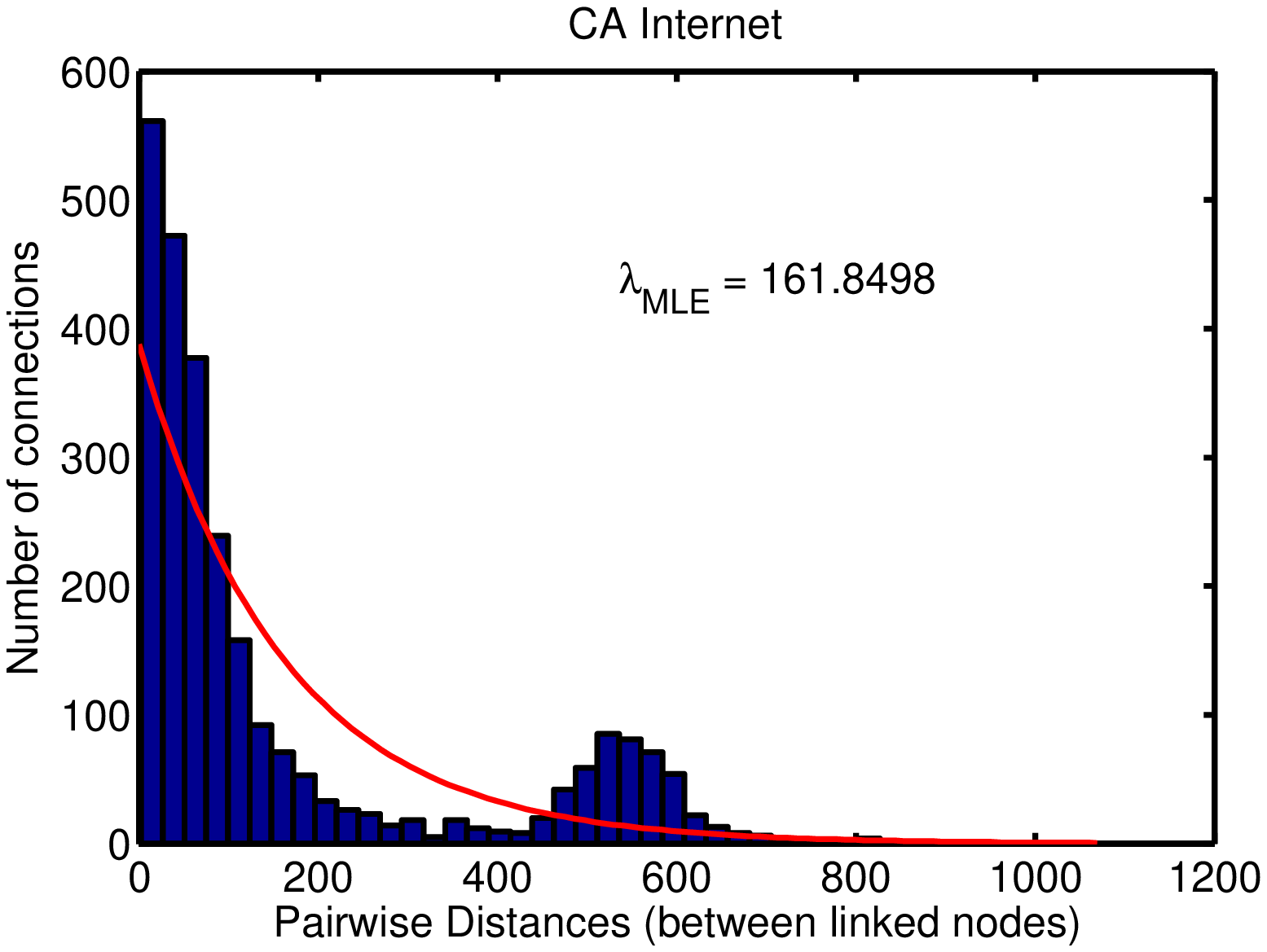}
 	}
 	\subfigure[US Airline]{
 	    \label{fig:dist_airline}
 	    \centering
 	    \includegraphics[keepaspectratio, width=0.23\textwidth]{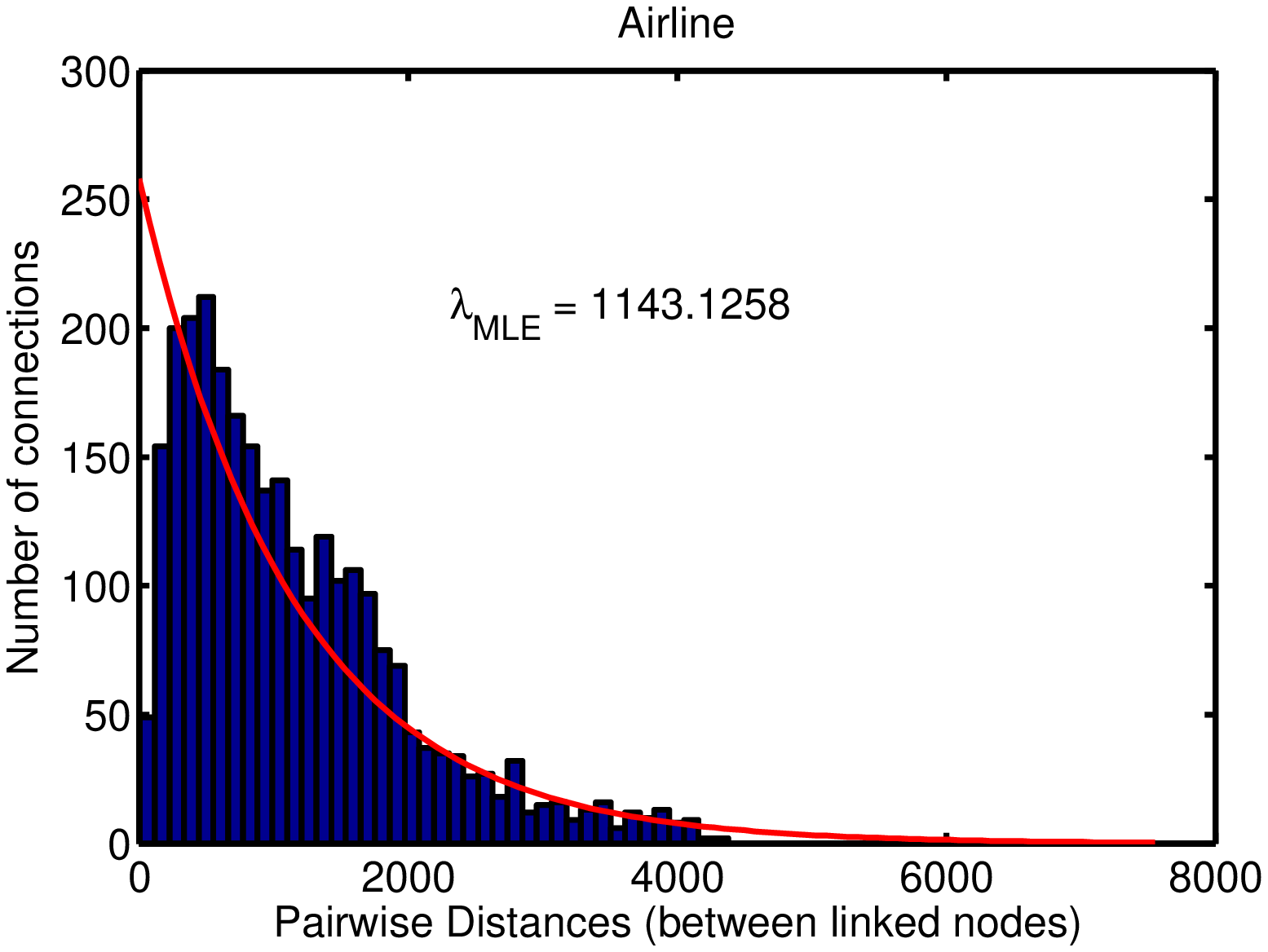}
 	}
 	\caption{Distribution of the pairwise distances between linked nodes along with a maximum likelihood fit to an exponential distribution.}
   \label{fig:distances}
\end{figure*}

We test the fit of an exponential distance decay function on four real-world spatial networks: \textit{C. elegans} neuron connections, social connections between users in Gowalla (a social photo sharing service), Internet server connections within California, and an airline transportation network for the United States (details provided in table~\ref{tbl:datasets}). We show the distribution of the pairwise distances of connected nodes in figure~\ref{fig:distances}, as well as a maximum likelihood fit to an exponential distribution. Although we see that only the Gowalla network potentially fits well to an exponential distribution, we perform a Kolmogorov-Smirnov (KS) test on each network to quantitatively test the fit. In fact, all of the networks reject the null hypothesis (that the data come from the same distribution) with p-values $4.6e^{-152}$ (\textit{C. elegans}), $2.2e^{-6}$ (Gowalla), $1.7e^{-55}$ (CA Internet), and $5.2e^{-29}$ (US Airline).

Additionally, the \textit{C. elegans} and CA Internet networks contain a small second mode in the tail of the distribution, caused by areas of heavy spatial clustering of the nodes. This tight interaction between the spatial distribution of nodes and the likelihood of observing long-distance connections makes it difficult to describe the distance with a single function over the entire network. 

In this paper, we investigate the variable effects of space on \textit{individual} nodes and how this influences network topology. To model these effects we combine ideas from previous spatial network models~\cite{Expert2011, Yook2002} with latent parameter models~\cite{Hoff2002, Hoff2009}. We capture the spatial effects with a latent, node-specific radius parameter. Furthermore, we extend this idea further by adding a second node-specific latent variable which captures space-independent community structure. Our experiments show that our model achieves up to $35\%$ improvements over other methods in the task of link prediction (in terms of area under the ROC curve). Moreover, we see much larger improvements (up to $80\%$) when predicting links between nodes with low degrees, where many link prediction techniques fail. 

\section{Related Work}
\label{sec:related_work}

The development of mathematical models of network structure has played an important role in advancing the area of network science~\cite{Albert2000, Barabasi1999, Erdos1960, Newman2003, Watts1998, White2006}. In this section we review the relevant research work in the areas of spatial network models and analysis and statistical network models.

\subsection{Spatial Networks}
\label{sec:spatnet}


The existing work on modeling spatial networks can be split into three general types of models: (i) Waxman models, (ii) geometric models, and (iii) preferential attachment and scale free spatial models. Perhaps the earliest model to incorporate the pairwise distance between nodes into the probability of a link was the Waxman model~\cite{Waxman1988}. Specifically, the authors proposed that the probability of a link is proportional to $Be^{-d_{ij}/L}$, for some constant $B$ and scaling coefficient $L$. The Waxman model can be construed as the spatial equivalent of the Erdos--Renyi random graph model (ER)~\cite{Erdos1960} since as $L \rightarrow \infty$, the model converges to the ER attachment model. While this spatial model has been shown to replicate some real world networks (e.g.~\cite{Kaiser2004a}), it fails to capture the preferential attachment that has been observed in many spatial and non--spatial networks.

The class of \textit{geometric} models, describe the probability of a link forming between two nodes as a function of distance which approaches one as the distance between two nodes decreases. Typically the probability of attachment is formulated as a logistic, $\frac{1}{1 + e^{-A(d_{ij} + B)}}$, where $A$ is a scale parameter controlling the slope of the logistic and $B$ controls the shift of the function. Pure geometric networks, where an edge between two nodes exists if the distance is less a certain threshold, can be considered a special case of a logistic function with $A \rightarrow \infty$. Many works have studied the theoretical network statistics of these thresholded graphs under the assumption of uniform spatial distribution \cite{Dall2002, Penrose2003}. Additionally, Wong et. al. \cite{Wong2006} propose a similar logistic spatial model for social networks that replicates several statistic of real world networks.

Traditional preferential attachment and scale--free network models have also been adapted to incorporate spatial information. Typically, the probability of attachment in these networks is proportional to $k_i e^{-d_{ij}/L}$ or $k_i d_{ij}^{A}$, such that one gets a network with preferential attachment that decays as an exponential or power law with distance~\cite{Yook2002}. Properties of these networks have been well studied~\cite{Barthelemy2003, Kosmidis2008, Barrat2005}, particularly that as $L$ and $A$ vary, the structure of the spatial networks can change from scale--free networks with little clustering to large networks with intense clustering~\cite{Barthelemy2003}. While these models are adept at modeling the evolution of complex spatial networks such as the Internet~\cite{Xulvi2002}, they still assume a homogeneous spatial effect throughout the network.

In addition to modeling, several authors have studied the structural properties of spatial networks and understand the role that space plays in the network topology. Specifically, there has been a large amount of work merging traditional network models with spatial models, and determining how these network models change under spatial constraints~\cite{Barthelemy2003, Barthelemy2011, Barnett2007, Daraganova2011, Ferretti2011, Hayashi2006, Turgut2010}. For instance, in~\cite{Hayashi2006}, the authors discuss how scale--free networks can be analyzed in a geometric space. The resulting models can be applied to several types of data to analyze the structural properties and provide insight into the link creation process. Such analyses are especially important in understanding biological networks~\cite{Kaiser2004a, Kaiser2004b}.

The distribution of nodes in space also affects the types of connections, and therefore the global structural properties of a spatial network. Bullock et al.~\cite{Bullock2010} discuss several properties of spatial networks and how the spatial distribution of the nodes effect these properties. For instance, when nodes are distributed uniformly in a given space, there is a sharp phase transition in the size of the largest component of the network, whereas nodes distributed in an inhomogeneous manner, exhibit a smooth transition in the number of connected components and their sizes. Additionally, Voges et al.~\cite{Voges2007} study the network properties (e.g. degree correlation, shortest path length, cluster coefficient, and spatial concentration) of networks embedded into a lattice. The authors experimented by adding some jitter to the node positions and studying the resulting of network statistics. They found that these properties are very sensitive to the randomness of the node locations. This further corroborates the importance of including the spatial properties of networks when studying their structural properties.

Beyond analyzing the structure of spatial networks, recent approaches to community detection in spatial networks propose new null network models, based on gravity models~\cite{Wilson1967}, which are implemented within the modularity framework~\cite{Newman2006}. The idea is to incorporate the pairwise distance between nodes into the expectation of whether or not a link exists between them, thus more accurately representing the spatial network structure~\cite{Cerina2012, Expert2011}. In Cerina et al.~\cite{Cerina2012}, the authors propose a model in which the probability of a link forming between two nodes declines exponentially as the distance between them increases. In Expert et al.~\cite{Expert2011}, the authors build an empirical distribution of the probability of connection conditioned on the distance from the observed network and use that to weight the connection probability. In both cases, the authors assume that the effect of distance remains constant throughout the entire network. Both of these models have shown to improve community findings in spatial networks over the originally proposed null model of preferential attachment (i.e. $\frac{k_i k_j}{2 \sum_t k_t}$).

In addition to descriptive modeling, Lennartsson et al.~\cite{Lennartsson2012} introduce \textit{SpecNet}, a general spatial network model that is capable of \textit{generating} networks with a full range of values for clustering coefficient, degree assortativity~\cite{Newman2002}, and fragmentation index. Whereas previous models were only able to create networks with a very limited range of possible statistics, \textit{SpecNet} is able to produce networks that can nearly cover the range of possible theoretical values for such measures. Such generative models provide a more concrete link between the various components of the network and how these relate to the structural properties.


\subsection{Latent Parameter Network Models}
\label{sec:statmodels}

Hoff et al.~\cite{Hoff2002} introduce a latent space approach for modeling social networks. The authors construct a model in which the objective is to infer node positions in a \textit{latent social space} such that links are more likely between nodes that are close together in this latent space. In fact, given each nodes' location in this latent social space, all of the network links are conditionally independent. This model is able to effectively represent a large number of social networks due to its ability to capture homophily. That is, nodes close together in latent space typically have similar distances to other nodes as well. Others have introduced interesting theoretical properties of this model as well as offered their own extensions~\cite{Handcock2007, Krivitsky2009, Sarkar2010}

Additionally, Hoff et al.~\cite{Hoff2005, Hoff2007, Hoff2009} have further developed more general latent factor models which have been shown to generalize~\cite{Hoff2002}. In~\cite{Hoff2007, Hoff2009}, the basic idea is to model network connections as $y_{i,j} \propto \beta X + uDu^{T}$, such that each link is a function of a set of covariates as well as a low rank approximation of node-wise random effects. The authors show that this model weakly generalizes the latent space and class models previously proposed, and provides high quality predictions for a wide variety of networks (e.g. social networks, word relationship networks, and protein interactions). In contrast, our objective in this work is to separate the set of dependent variables such that we isolate the \textit{spatial term} from the others. As our hypothesis is that spatial effects vary over the network, we want to study the effect on each node in the original space.

Lastly, block models are another form of latent variable models, often used for community detection, in which each node is associated with a latent group parameter such that nodes are more likely to form connections within a group than between groups~\cite{Airoldi2008, Karrer2011}. These models assume nodes fall into equivalence classes such that the probability of a pair of nodes connecting is conditionally independent given the latent group identifiers of nodes. The inferential problem is then to compute the latent class identifier for each node, given the network structure. For a more comprehensive survey of the work in this area, we refer the reader to~\cite{Airoldi2009}.



\section{Node-Centric Spatial Network Model}
\label{sec:model}

In this section we introduce a novel probabilistic model for analyzing spatial networks in which spatial effects are captured at the level of individual nodes. To capture the variable effects of space throughout the network, we introduce a latent, positive real-valued, parameter referred to as the radius at each node. We introduce two models which incorporate this idea, \textit{Radius} and \textit{Radius+Comms}. The first model, \textit{Radius}, only models the node-specific spatial effects and node popularity. The second model, \textit{Radius+Comms}, adds a component to capture community structure within the network which cannot be explained by factors incorporated in the \textit{Radius} model.


Throughout this work, we assume that we are given as input a spatial network. A network is represented by the adjacency matrix, $A$, where $A_{ij} = 1$ if there is a link between nodes $i$ and $j$. The degree of a node is computed by summing over a particular row of $A$, $k_i = \sum_j A_{ij}$. The pairwise distances between nodes is given by the matrix, $D$, such that $D_{ij}$ is the Euclidean distance between nodes $z_i$ and $z_j$.

\subsection{Basic Spatial Model: \textit{Radius}}
\label{sec:basicmodel}

The \textit{Radius} model is based on the idea that space may influence each node differently. The model consists of two terms, (i) a spatial term which favors forming links between nodes in which their radius-corrected pairwise distance is small and (ii) a preferential attachment term which favors forming links between nodes with high degrees. We combine both of these terms within the logistic function since the output is interpreted as the probability of an edge existing between two nodes. The probability of forming a link is defined in Eq.~\ref{equ:likelihood}.
\begin{eqnarray}
  \label{equ:likelihood}
	p(A | R, D, K, \alpha, \gamma) & = & \frac{1}{1 + exp\left(-\frac{1}{\alpha}(r_i + r_j - D_{ij}) + \frac{1}{\gamma} \left(\frac{k_i k_j}{\sum_z k_z} - M \right) \right)}
\end{eqnarray}
The first term, $\frac{1}{\alpha}(r_i + r_j - D_{ij})$, describes the propensity of a pair of nodes to form a link given their (latent) radius parameters and the distance separating them. Although it is more costly to form long distance links in general, the radii can reduce or even completely overcome this cost. The scale parameter, $\alpha$, controls the strength of the distance term on the overall link probability. This parameter also allows the model to automatically adapt to networks at different scales.

\begin{figure}[h]
    \centering
    \includegraphics[keepaspectratio, width=0.45\textwidth]{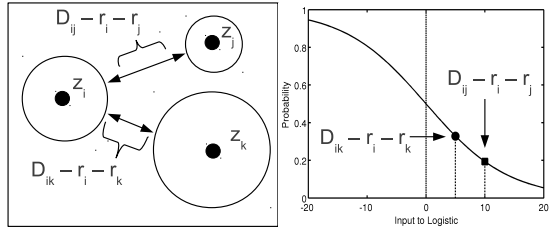}
    \caption{Illustration of how the radii from different nodes interact with each other and the pairwise distance to determine the existence of an edge.}
    \label{fig:radius_conn}
\end{figure}
Figure~\ref{fig:radius_conn} illustrates the role of the radii in forming a link between two nodes separated by distance, $D_{ij}$. Although nodes may be separated by a large distance, if the combined radii can make up for this distance, or at least reduce it, a link between these nodes becomes more likely. That is, we assume a simple linear relationship between radii and pairwise distance: $D_{ij} - r_i + r_j$. Since we would like to predict the output of $0/1$, depending on whether an edge exists or not, we place this term into a logistic function.

The second term describes the propensity of nodes to form links with popular nodes (i.e. nodes with a large degree). This is the standard term considered in preferential attachment-based models of network structure. The constant $M$ is the midpoint between the average combined degree of the set of nodes for which a link exists and the average combined degree of the set of nodes for which a link doe not exist. That is, if $k_x k_y < M < k_i k_j$, then, given no other information, $p(A_{ij}) > p(A_{xy})$. Including this constant allows this term, $\frac{k_i k_j}{\sum_z k_z} - M$, to take on both positive and negative values. Since it is placed into a logistic function, this allows us to both increase and decrease the overall probability of a link. The parameter, $\gamma$, is again a scaling parameter which controls the total influence of this term on the resulting link. The two scaling parameters offer a large degree of flexibility to the model since it is able to automatically adapt to networks with both very strong and very weak spatial effects.

The posterior distribution for our model is given in Eq.~\ref{equ:post_base}. Our objective is to infer values of the hidden variables, $\alpha$, $\gamma$, and $R$ (the vector of radii), given the observed network structure, $A$, node degrees, $K$, and pairwise distances, $D$. We use truncated Gaussian distributions, denoted $\mathcal{N}_{>0}()$, for priors over all of the latent variables in our model (since all of the variables are restricted to be positive). We discuss the inference computation more in section~\ref{sec:inference}.

\begin{eqnarray}
    \label{equ:post_base}
    p(R, \alpha, \gamma | A, K, D) & \propto & p(A | R, D, K, \alpha, \gamma) p(R) p(\alpha) p(\gamma) \nonumber \\
				   & = & p(\alpha) p(\gamma) \prod_{i > j}^{n} p(A_{ij} | r_i, r_j, D_{ij}, k_i, k_j, \alpha, \gamma) p(r_i) p(r_j) \nonumber \\
  				   & = & \mathcal{N}_{>0}(\alpha; \mu_{\alpha}, \sigma_{\alpha}) \; \mathcal{N}_{>0}(\gamma; \mu_{\gamma}, \sigma_{\gamma}) \nonumber \\
				   & \times & \prod_{i > j}^{n} logistic \left(\frac{1}{\alpha}((r_i + r_j) - D_{ij}) + \frac{1}{\gamma} \left(\frac{k_i k_j}{\sum_z k_z} - M \right) \right) \nonumber \\
				    & \times & \mathcal{N}_{>0}(r_i; \mu_{r}, \sigma_{r}) \; \mathcal{N}_{>0}(r_j; \mu_{r}, \sigma_{r})
\end{eqnarray}

\subsection{Community Model: \textit{Radius+Comms}}
\label{sec:communitymodel}

\begin{figure}[h]
    \centering
    \includegraphics[keepaspectratio, width=0.7\textwidth]{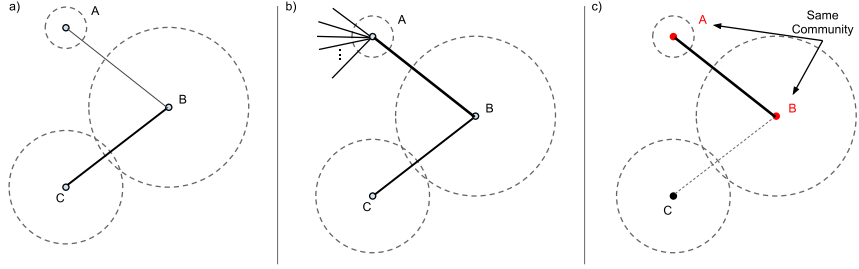}
    \caption{The different mechanisms that may influence the probability of a connection between two nodes. In each of the instances, the distance from node $A$ to $B$ and from node $C$ to $B$ are equal. In figure (a) the link probabilities are determined by the combined radii of the nodes. It is much more likely that nodes $B$ and $C$ will form a link due to their radii. In figure (b), the probably of a link between nodes $A$ and $B$ increases because node $A$ is a hub (i.e. high node degree), even though it still has a small spatial reach. In figure (c), nodes $A$ and $B$ have a high probability of forming a link because they are both in the same community. In contrast the probability of a connection between $B$ and $C$ is reduced because they are in different communities.}
    \label{fig:connection_examples}
\end{figure}

Although nodes that are physically close together are more likely to form a link than nodes that are further apart, space is not the only factor in deciding which nodes should be connected. Previous literature~\cite{Airoldi2009, Barthelemy2011} often identify three main explanations of links: (i) close spatial proximity, (ii) node popularity, and (iii) community structure within the network. These factors are illustrated in figure~\ref{fig:connection_examples}.

With the basic model in place, we develop a simple extension, \textit{Radius+Comms}, which allows us to simultaneously infer any space-independent community structure within the network as well. To describe the community structure, we attach a discrete latent parameter to each node which identifies the node's group label, $c_i \in \{0, ..., K\}$. Nodes within the same community should have more links to other nodes within their community and fewer links to nodes in other communities. We model this by adding a (latent) random variable within the logistic function. This way the community effects do not completely override spatial behavior of nodes, rather they can strengthen or dampen the effects of distance on a particular connection to make it a more probable outcome.

Unlike most community detection methods, we offer a \textit{don't care} community ($c_i = 0$) which allows the formation of links between nodes to follow only the previously described model. That is, for nodes placed into the \textit{don't care} community, the probability of a link involving this node remains unchanged, even if the link connects to a node in another community. This formulation ensures that our model will only capture salient network structure which cannot otherwise be explained by other factors. The new community term, $\beta(c_i, c_j)$, is given in Eq.~\ref{equ:beta}.
\begin{eqnarray}
  \label{equ:beta}
  \beta(c_i, c_j) =  \left\{\begin{array}{l l}
			    0 & \quad c_i = 0 \; \text{ or } \; c_j = 0 \\
			    \phi & \quad c_i = c_j\\
			    -\phi & \quad c_i \ne c_j\\
			    \end{array} \right.
\end{eqnarray}
If nodes belong to the same community, we increase the probability of a connection by adding $\phi$ to the other terms within the logistic function. Where $\phi$ is a positive, real-valued random variable to be inferred from the observed data. Combining this with our previous model, the updated posterior distribution is given in Eq.~\ref{equ:post_comm}.
\begin{eqnarray}
    \label{equ:post_comm}
    p(R, C, \alpha, \gamma, \phi | A, K, D) & \propto & \mathcal{N}_{>0}(\alpha; \mu_{\alpha}, \sigma_{\alpha}) \; \mathcal{N}_{>0}(\gamma; \mu_{\gamma}, \sigma_{\gamma}) \;
							\mathcal{N}_{>0}(\phi; \mu_{\phi}, \sigma_{\phi}) \nonumber \\
				& \times & \prod_{i > j}^{n} logistic \left(\frac{1}{\alpha}((r_i + r_j) - D_{ij}) + \beta(c_i, c_j) + \gamma \left(\frac{k_i k_j}{\sum_z k_z} - M \right) \right) \nonumber \\
				& \times &  \mathcal{N}_{>0}(r_i; \mu_{r}, \sigma_{r}) \; \mathcal{N}_{>0}(r_j; \mu_{r}, \sigma_{r}) \nonumber \\
				& \times & \text{multinomial}(c_i; \theta_C) \; \text{multinomial}(c_j; \theta_C)
\end{eqnarray}
The new random vector, $C$, encodes the community IDs for each node and if $c_i = 0$, then this node is assigned to the \textit{don't care} community. The interaction between nodes within the same community and across communities is modified by the function $\beta(c_i, c_j)$ which is defined in Eq.~\ref{equ:beta}. This adds one extra weighting (positive, real-valued) variable, $\phi$. If $c_i = c_j$, then a large value of $\phi$ will increase the probability of a link between the two nodes, whereas if $c_i \ne c_j$, then $-\phi$ will decrease the probability of a link. Note that this defines a symmetric relationship; within-group connections are strengthened by the same amount that between-group connections are penalized.

The number of clusters, $K$, should be set sufficiently large to accommodate any structure that may exist. Because we include a \textit{don't care} community, the specific setting of $K$ is not critical since, if there is insufficient evidence of clustering, nodes may simply be assigned $c_i = 0$. However, as $K$ increases, the rate of convergence of our inference routine may slow, since it much search a larger discrete space. In our experiments, we set $K$ to $10\%$ of the number of nodes in the network. We have found that this provides a nice trade-off between flexibility and efficiency as confirmed by our analysis of the MCMC trace plots. In fact, many of the networks we have tested identify fewer communities, and only the \textit{C. elegans} network places every node into a community.

\subsection{Inference}
\label{sec:inference}

To compute with our model, we employ a standard Markov Chain Monte Carlo (MCMC) algorithm for approximate inference. We chose to apply Bayesian inference rather than maximum likelihood or stochastic search optimization to ensure that all of the uncertainty was appropriately propagated throughout the model. Just as it is unlikely that there exists a single global function over distance which can accurately capture the effects over the whole network, we do not expect the inferred radius values to be exact measures of the nodes' spatial reach.
\begin{algorithm}[h]
    \caption{Metropolis within Gibbs sampling routine for Bayesian inference of our spatial network model.}
    \label{alg:mcmc}
    {\footnotesize
	\begin{algorithmic}
	    \STATE // Randomly initialize random variables: $\alpha, \gamma, r_i \forall i$
	    \FOR{s = $1 \to$ T}
		\STATE // propose new values for global variables
		\STATE $\hat{\alpha} \sim \mathcal{N}(\alpha_{s-1}, \sigma_{\alpha})$, $\hat{\gamma} \sim \mathcal{N}(\gamma_{s-1}, \sigma_{\gamma})$
		\STATE // compute acceptance ratio
		\STATE acceptRatio $= (logP(A | R, D, \hat{\alpha}, \hat{\gamma}) + logP(\hat{\alpha}) + logP(\hat{\gamma})) - (logP(A | R, D, \alpha^{s-1}, \gamma^{s-1}) + logP(\alpha^{s-1}) + logP(\gamma^{s-1}))$
		\STATE $u \sim$ unif(0, 1)
		\IF{$log(u) <$ acceptRatio}
			\STATE $\alpha^{s} = \hat{\alpha}$, $\gamma^{s} = \hat{\gamma}$ // accept samples
		\ELSE
			\STATE $\alpha^{s} = \alpha^{s-1}$, $\gamma^{s} = \gamma^{s-1}$ // reject samples
		\ENDIF
		\STATE // propose new values for node variables
		\FOR{$j = 1 \to n$}
			\STATE $\hat{r_i} \sim \mathcal{N}(r_i^{s-1}, \sigma_{r})$
			\STATE acceptRatio $= (logP(A_{i-} | R_{-i}, \hat{r_i}, D, \alpha, \gamma) + logP(\hat{r_i})) - (logP(A_{i-} | R_{-i}, r_i^{s-1}, D, \alpha, \gamma) + logP(r_i^{s-1}))$
			\STATE $u \sim$ unif(0, 1)
			\IF{$log(u) <$ acceptRatio}
				\STATE $r_i^{s} = \hat{r_i}$ // accept sample
			\ELSE
				\STATE $r_i^{s} = r_i^{s-1}$ // reject sample
			\ENDIF
		\ENDFOR
	\ENDFOR
\end{algorithmic}}
\end{algorithm}

The sampling procedure iterates between proposing new global parameter values (i.e. scaling parameters) with new radius values. Algorithm~\ref{alg:mcmc} outlines the full MCMC algorithm for the \textit{Radius} model. Inference on \textit{Radius+Comms} is a straightforward extension of this algorithm where we also infer the value of $\phi$, the global community penalty and reward as well as the $c_i$'s, the group ID's for each node. 

We use the notation $logP$ to refer to the log of the probability density function. The vector, $R$, is the set of all radii, whereas $R_{-i}$ is all of the radiis except for $r_i$. We use truncated Gaussians for all of the prior distributions since all of the parameters are restricted to positive values. Additionally, we set the parameters for the prior distributions to be rather uninformative, though specific to each network due to the differences in distance scales across our datasets. Lastly, we have experimented with different block-updating schemes, however, the one presented here, in which we first update the global scaling parameters, then each of the node parameters provided relatively fast convergence and good mixing for all of the networks (more discussion on this in section~\ref{sec:mcmc}.

\section{Experiments}
\label{sec:experiments}

We experimentally evaluate our proposed model by applying it to the task of link prediction on four different real-world spatial networks (described in table~\ref{tbl:datasets}). Furthermore, we offer additional analysis of the model parameters and present interesting interpretations by utilizing additional information about the network nodes.

\subsection{Analysis of Inferred Radii}
\label{sec:radius}
\begin{figure*}[ht]
    \centering
    \subfigure[\textit{C. elegans} (\textit{Radius})]{
	\label{fig:celegans_rvk}
	\centering
	\includegraphics[keepaspectratio, width=0.22\textwidth]{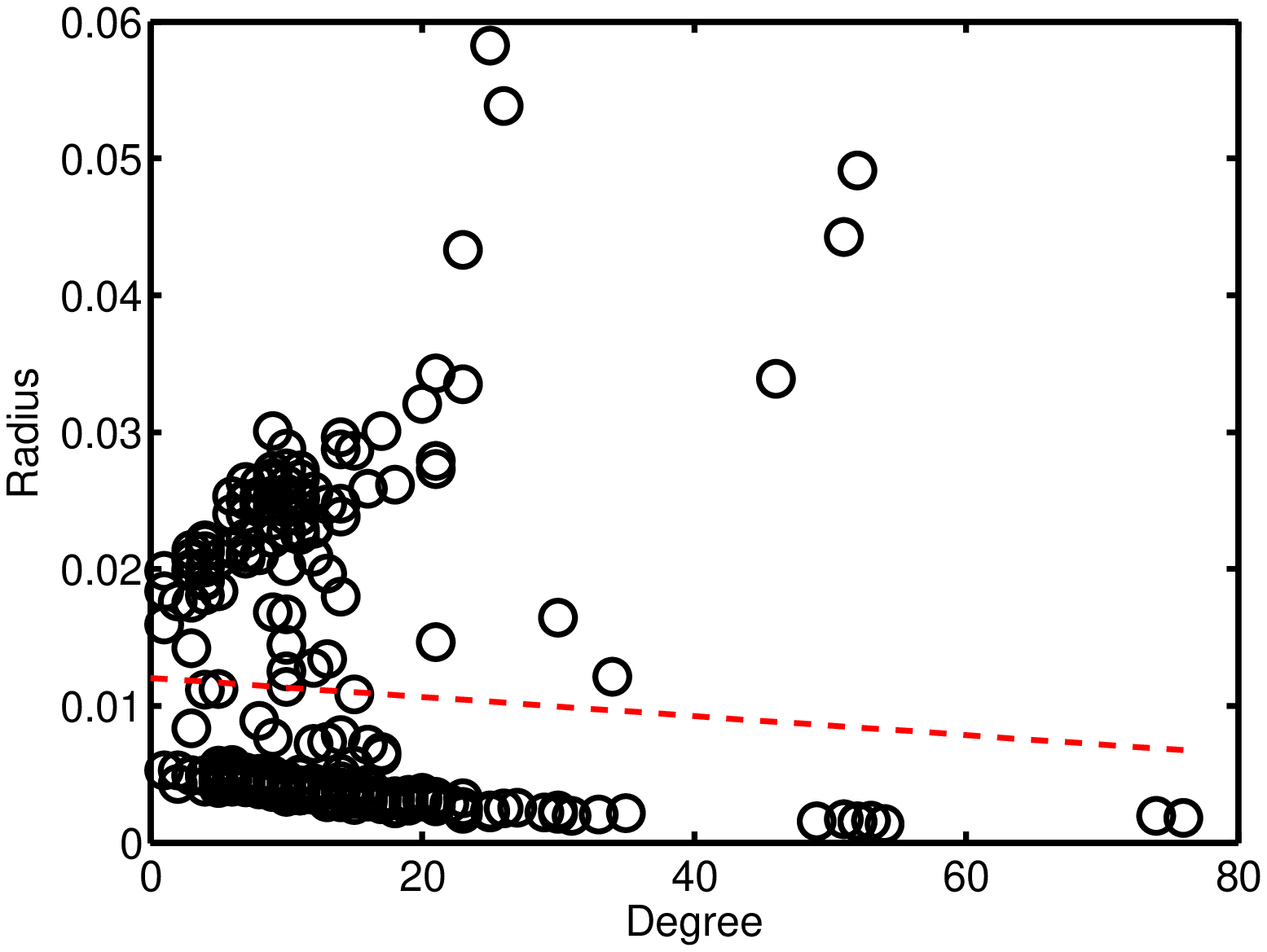}
    }
    \subfigure[Gowalla (\textit{Radius})]{
	\label{fig:gowalla_rvk}
	\centering
	\includegraphics[keepaspectratio, width=0.22\textwidth]{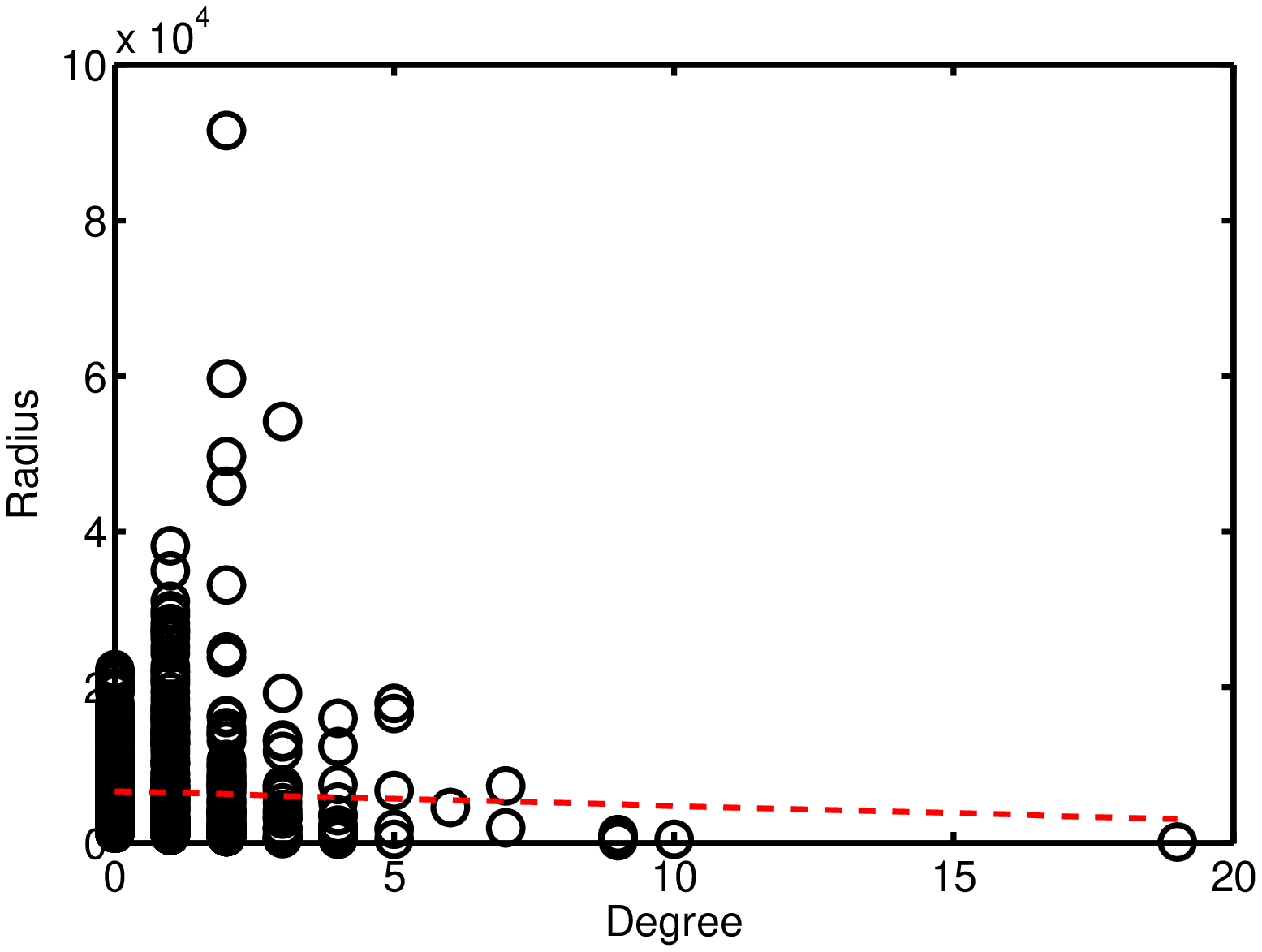}
    }
    \subfigure[CA Internet (\textit{Radius})]{
	\label{fig:internet_rvk}
	\centering
	\includegraphics[keepaspectratio, width=0.22\textwidth]{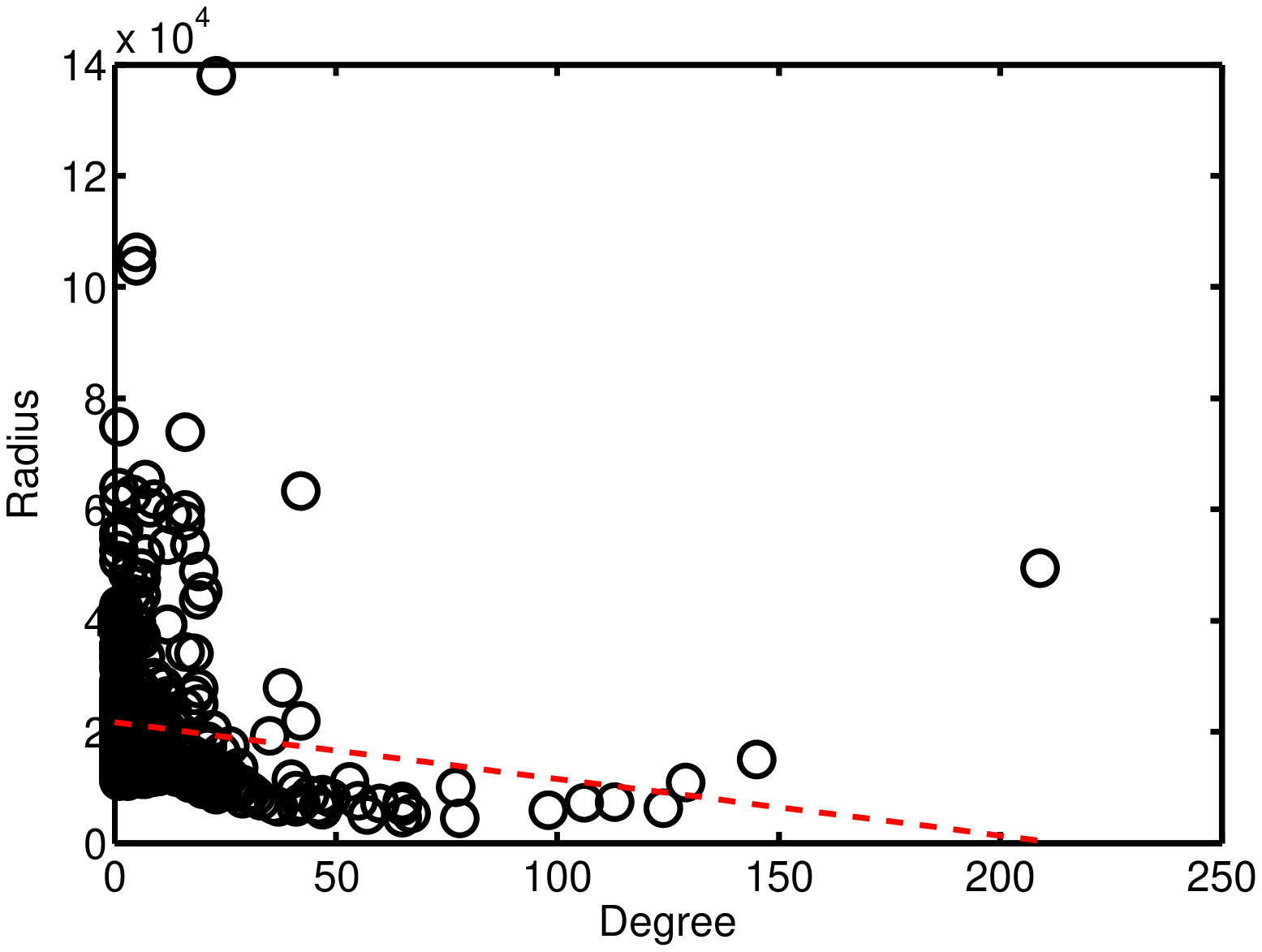}
    }
    \subfigure[Airline (\textit{Radius})]{
	\label{fig:airline_rvk}
	\centering
	\includegraphics[keepaspectratio, width=0.22\textwidth]{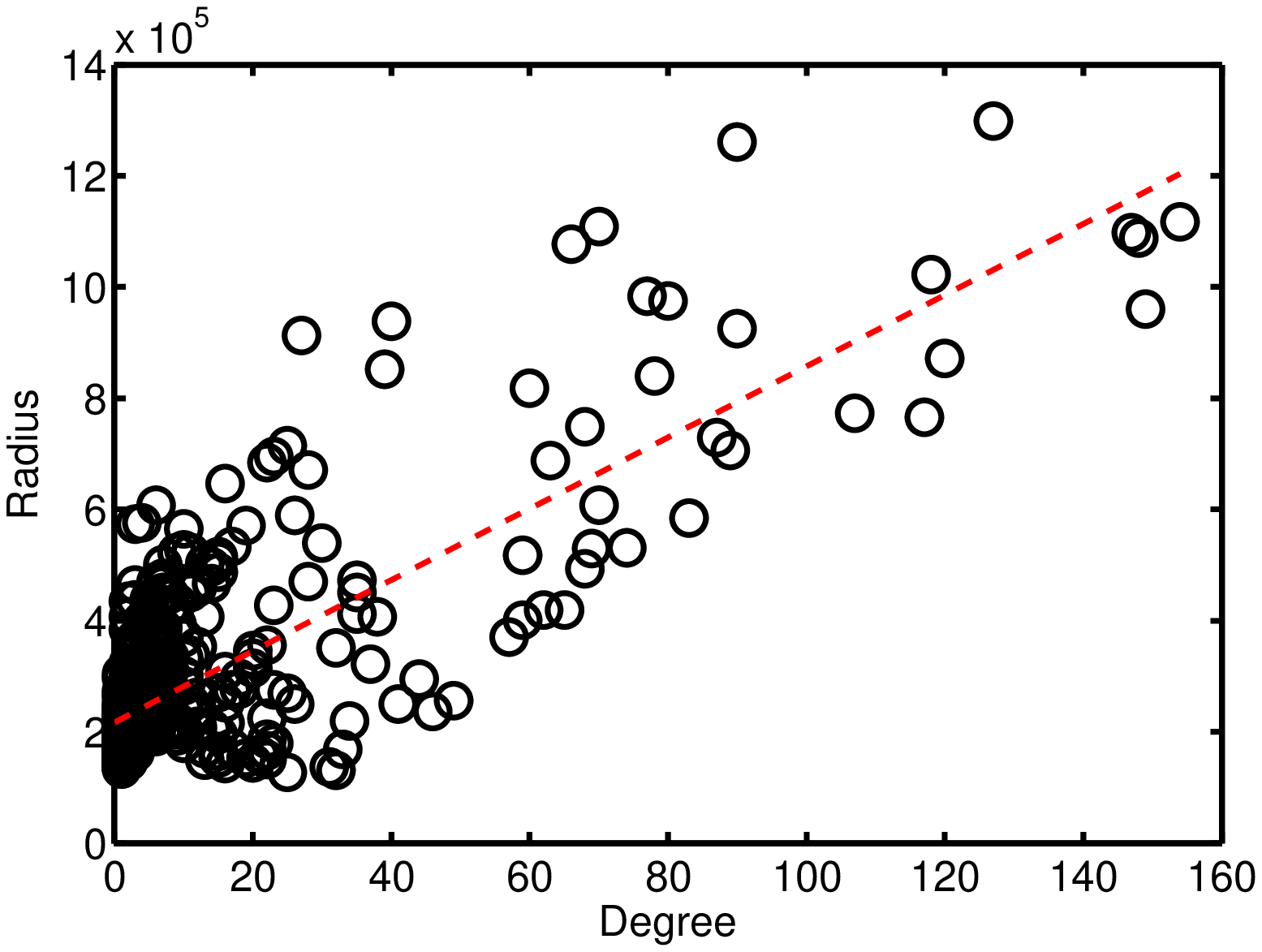}
    }

    \subfigure[\textit{C. elegans} (\textit{Radius+Comms})]{
	\label{fig:celegans_rvk1}
	\centering
	\includegraphics[keepaspectratio, width=0.22\textwidth]{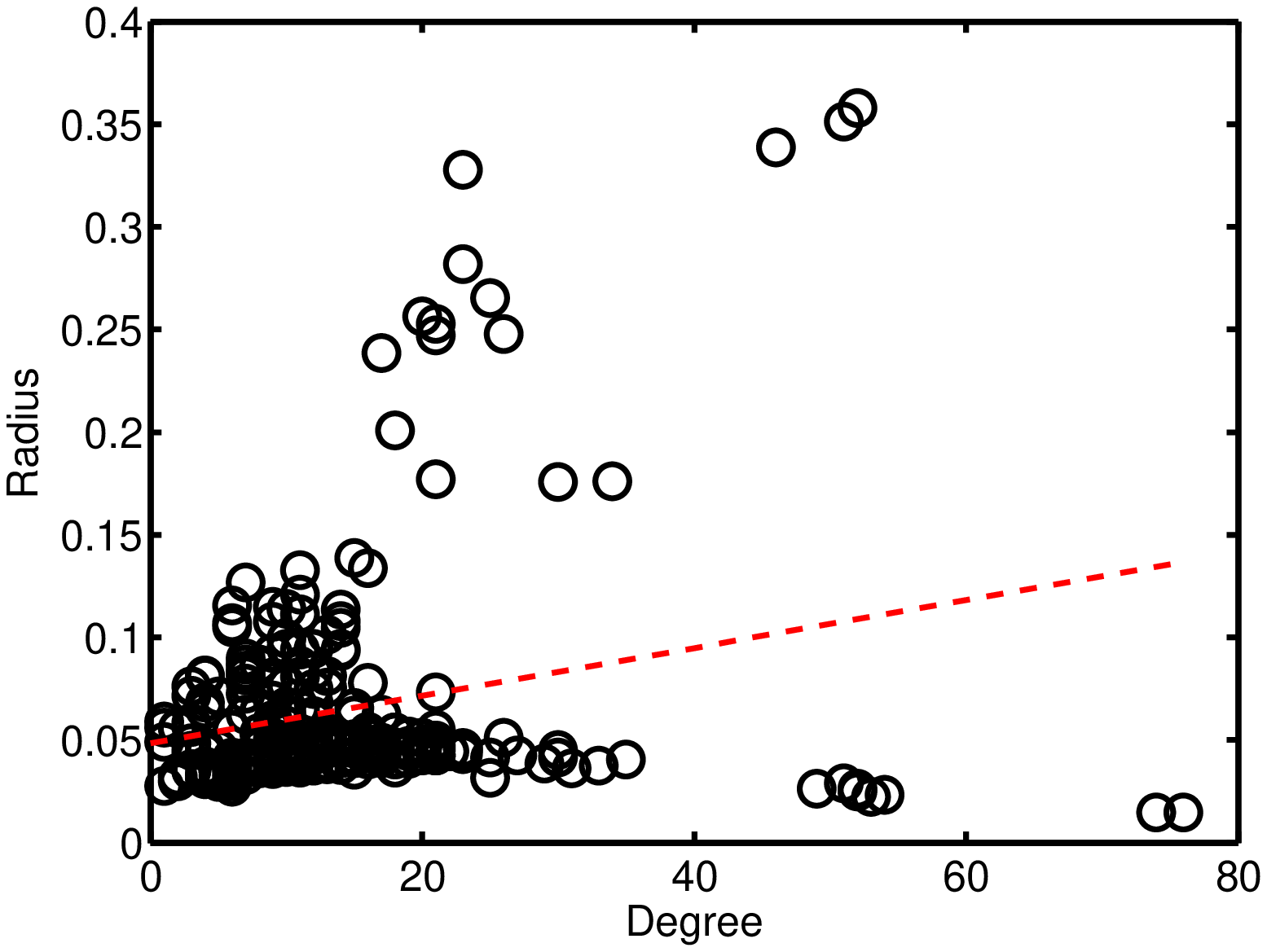}
    }
    \subfigure[Gowalla (\textit{Radius+Comms})]{
	\label{fig:gowalla_rvk1}
	\centering
	\includegraphics[keepaspectratio, width=0.22\textwidth]{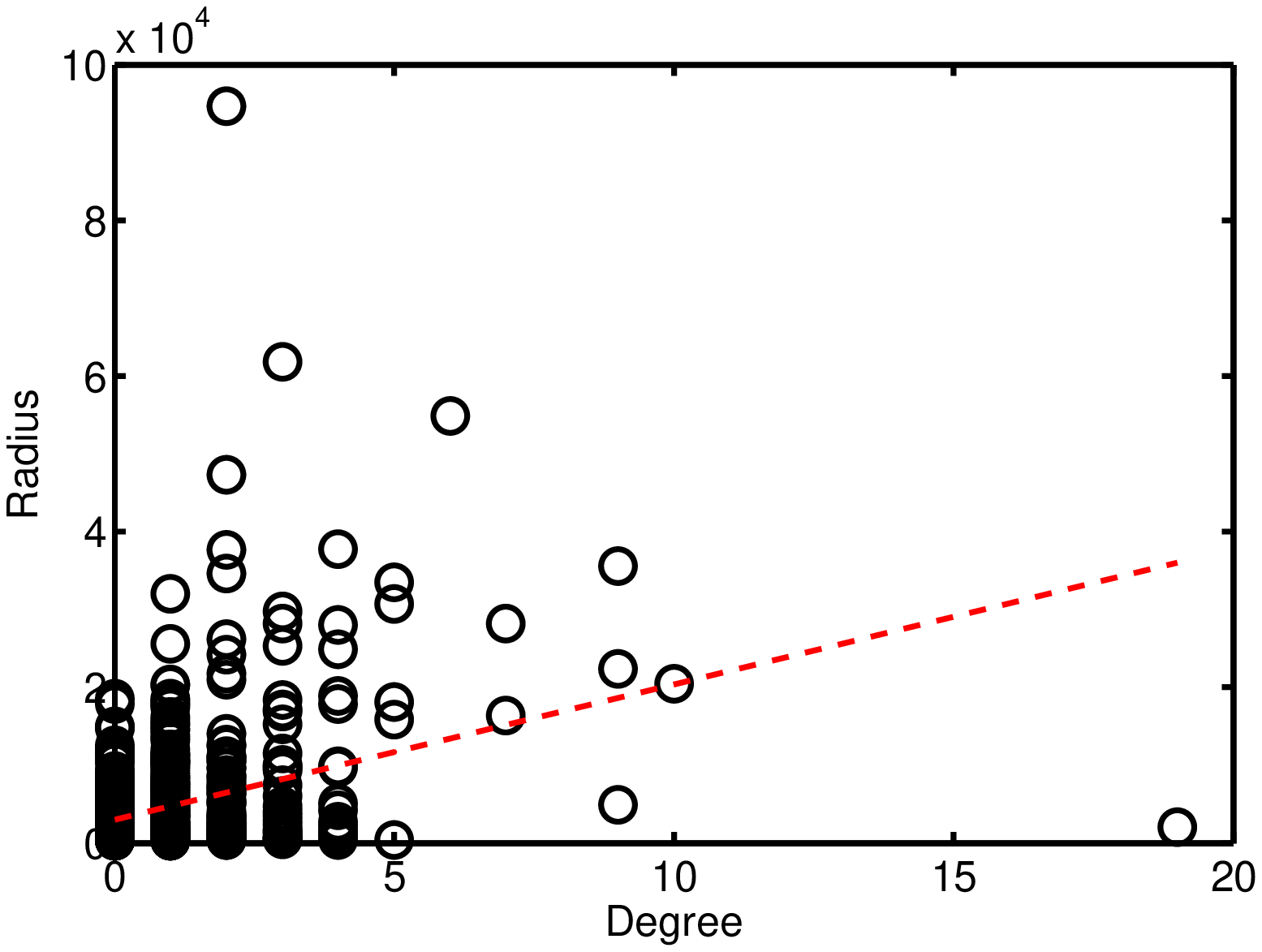}
    }
    \subfigure[CA Internet (\textit{Radius+Comms})]{
	\label{fig:internet_rvk1}
	\centering
	\includegraphics[keepaspectratio, width=0.22\textwidth]{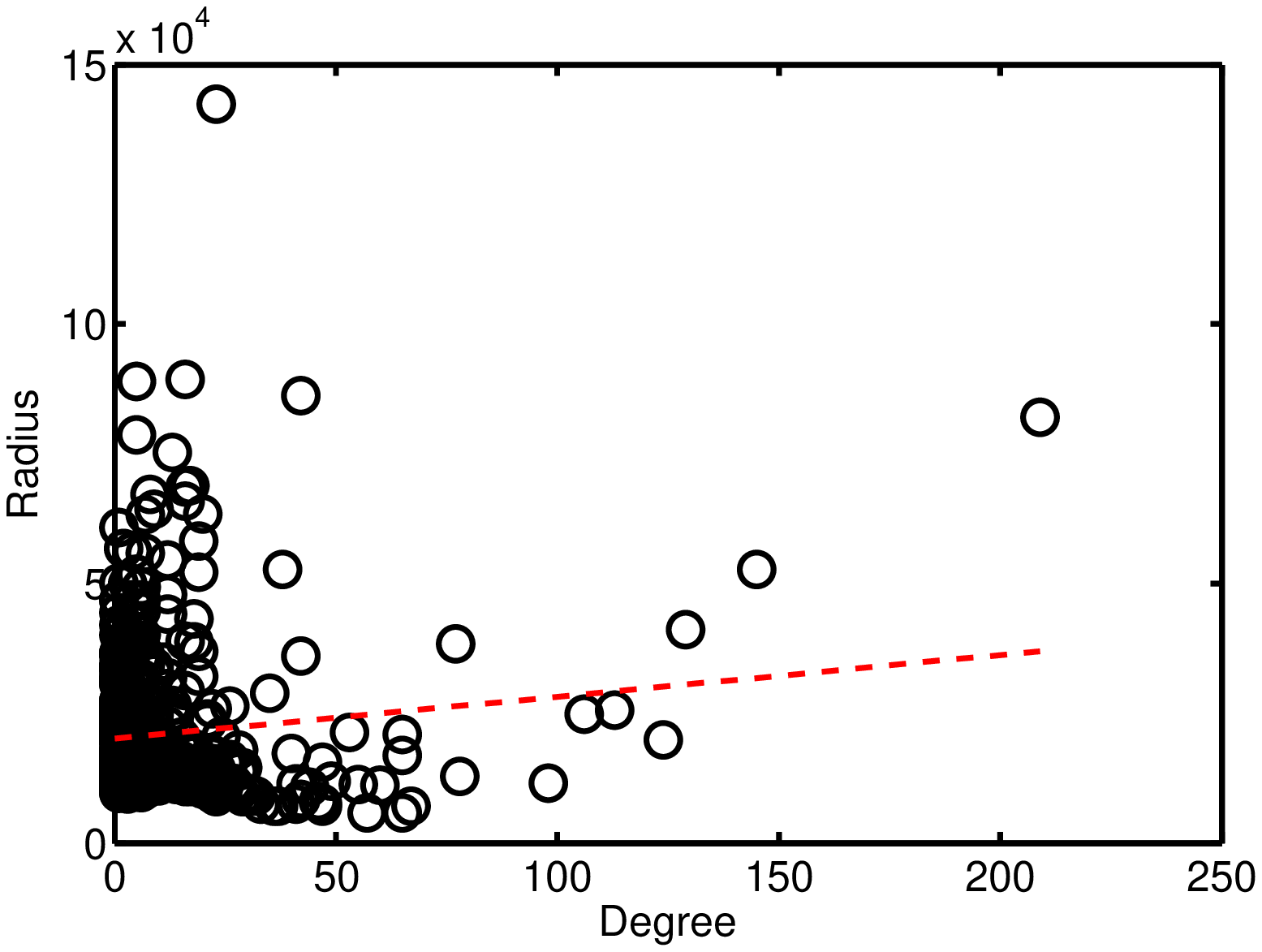}
    }
    \subfigure[Airline (\textit{Radius+Comms})]{
	\label{fig:airline_rvk1}
	\centering
	\includegraphics[keepaspectratio, width=0.22\textwidth]{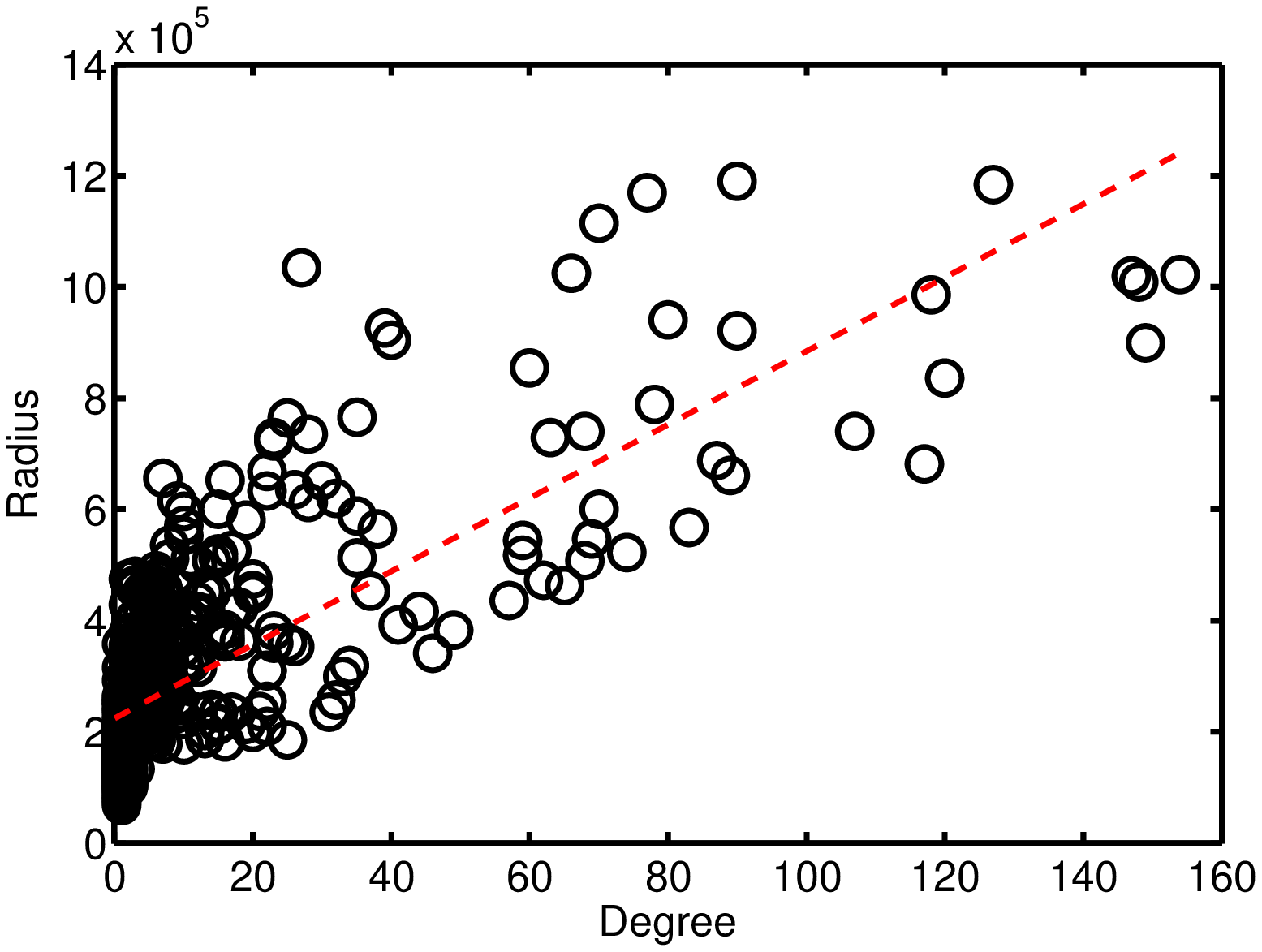}
    }
    \caption{Degree versus mean posterior radius for each network. The dotted line in each figure is the ordinary least squares regression fit to this data, where degree is the covariate and radius is the response (i.e. $radius = m \; degree + b$). The Pearson correlation between mean posterior radius and degree for the \textit{Radius} (\textit{Radius+Comms}) model for each network is \subref{fig:celegans_rvk} $-0.07 (0.23)$, \subref{fig:gowalla_rvk} $-0.03 (0.32)$, \subref{fig:internet_rvk} $-0.14 (0.11)$, and \subref{fig:airline_rvk} $0.78 (0.77)$.}
    \label{fig:rvk}
\end{figure*}

We have shown our model performs well on two common tasks, link prediction and community detection. Next, we investigate the inferred radii in more detail. Our claim was that the radius was meant to capture a node's spatial reach. While this is related to the degree of a node, we show that the radius will contain additional, unique information about a node's propensity to take part in long (short) distance connections. To test this, we plot the mean posterior radius for each node against its degree and test the amount of correlation in these values. We do this for both models and compare our results, shown in figure~\ref{fig:rvk}.

From figure~\ref{fig:rvk}, we make three interesting observations. First, there is a large variance in the inferred radii values corroborating our claim that distances effect individuals in a different manner. For example, in the \textit{C. elegans} network, we see clusters around different radii for nodes with similar degrees. This likely corresponds to the spatial clustering of neurons in both the head and the tail of the worm. Neurons in the head require a much smaller spatial reach since they have many potential connections within a short distance. Similarly, neurons in the tail also cluster spatially, however, to a lesser degree, thus requiring a slightly larger radius. We see a similar pattern in each of the networks, though to a lesser degree since connections in these networks are much more localized than in \textit{C. elegans}.

Second, there is little correlation between node degree and mean posterior radius. This indicates that the inferred radius values are capturing the spatial tendencies of each node, rather than simply re-capturing a measure of node popularity. In fact, only the Airline network shows any significant correlation between these two values. We also notice that this is the only network for which the nodes are distributed nearly uniformly at random (see \textit{index of dispersion} in table~\ref{tbl:datasets}). When nodes are uniformly distributed, there will be little difference in any node's spatial reach since all nodes must extend approximately the same distance in order to reach another node. Thus nodes which take part in more connections will tend to extend further.

Third, the distribution of radii is different for the two models with no clear trend across all networks. The additional modeling power in \textit{Radius+Comms} is used primarily to explain away the presence of abnormally long distance connections as well as the absence of closely co-located nodes of medium to high degree. In the first case, the radius for each of the nodes involved may be reduced since the abnormally long link is explained by an additional factor. In contrast, in the second case, the radii may grow larger, since the penalty of the two nodes belonging to different communities sufficiently explains why they do not connect. Depending on the particular network, we will likely see a mix of these two cases, thus causing some radii to grow and others to shrink accordingly.


\subsection{Link Prediction}
\label{sec:linkpred}

We first evaluate our model by performing link prediction using $10$-fold cross validation with a $90/10$ split for training and testing (i.e. $90\%$ of the links are used for training the model and the remaining $10\%$ are predicted) over each of the spatial networks. We compute the link predictions with our model in two different manners: (i) the predictive link probability and (ii) the maximum a-posterior (MAP) parameter configuration of the model. The predictive link probability, given in Eq.~\ref{equ:linkpred_pred}, is defined by integrating over the posterior probabilities of the model parameters to compute the probability of a link existing.
\begin{eqnarray}
  \label{equ:linkpred_pred}
  p(A_{ij} | D_{ij}, k_i, k_j) & = & \int_{\alpha, \gamma, r_i, r_j} p(A_{ij} r_i, r_j, \alpha, \gamma | D_{ij}, k_i, k_j) d\alpha d\gamma dr_i dr_j
\end{eqnarray}
Whereas using the MAP configuration simply requires plugging in the set of parameters that maximized the posterior probability. More formally, the MAP link prediction is given as follows:
\begin{eqnarray}
  \label{equ:linkpred_map}
  p(A_{ij} | D_{ij}, k_i, k_j) & = & p(A_{ij} | r_i^*, r_j^*, D_{ij}, k_i, k_j, \alpha^*, \gamma^*) \\
  \{r_i^*, r_j^*, \alpha^*, \gamma^*\} & = & argmax \; p(A_{ij} r_i, r_j, \alpha, \gamma | D_{ij}, k_i, k_j) \nonumber
\end{eqnarray}
Both of these methods consistently gave similar predictions, thus we only show results using the predictive link probability. To provide a baseline, we compare our model to (i) preferential attachment (PA), (ii) PA with exponential distance decay (ExpDist)~\cite{Cerina2012, Yook2002}, and (iii) PA with empirical distance decay (EmpDist)~\cite{Expert2011}. To perform link prediction using these methods, we compute the expectation of an edge for each pair of nodes using the statistics collected from the training links. Because the normalizations used in each of these methods is based on the total number of links in the network, the expectation may result in values larger than $1$. These values are thresholded and simply taken to be $1$.

\begin{figure*}[ht]
	\centering
	\subfigure[\textit{C. elegans}]{
	    \label{fig:celegans_auc}
	    \centering
	    \includegraphics[keepaspectratio, width=0.22\textwidth]{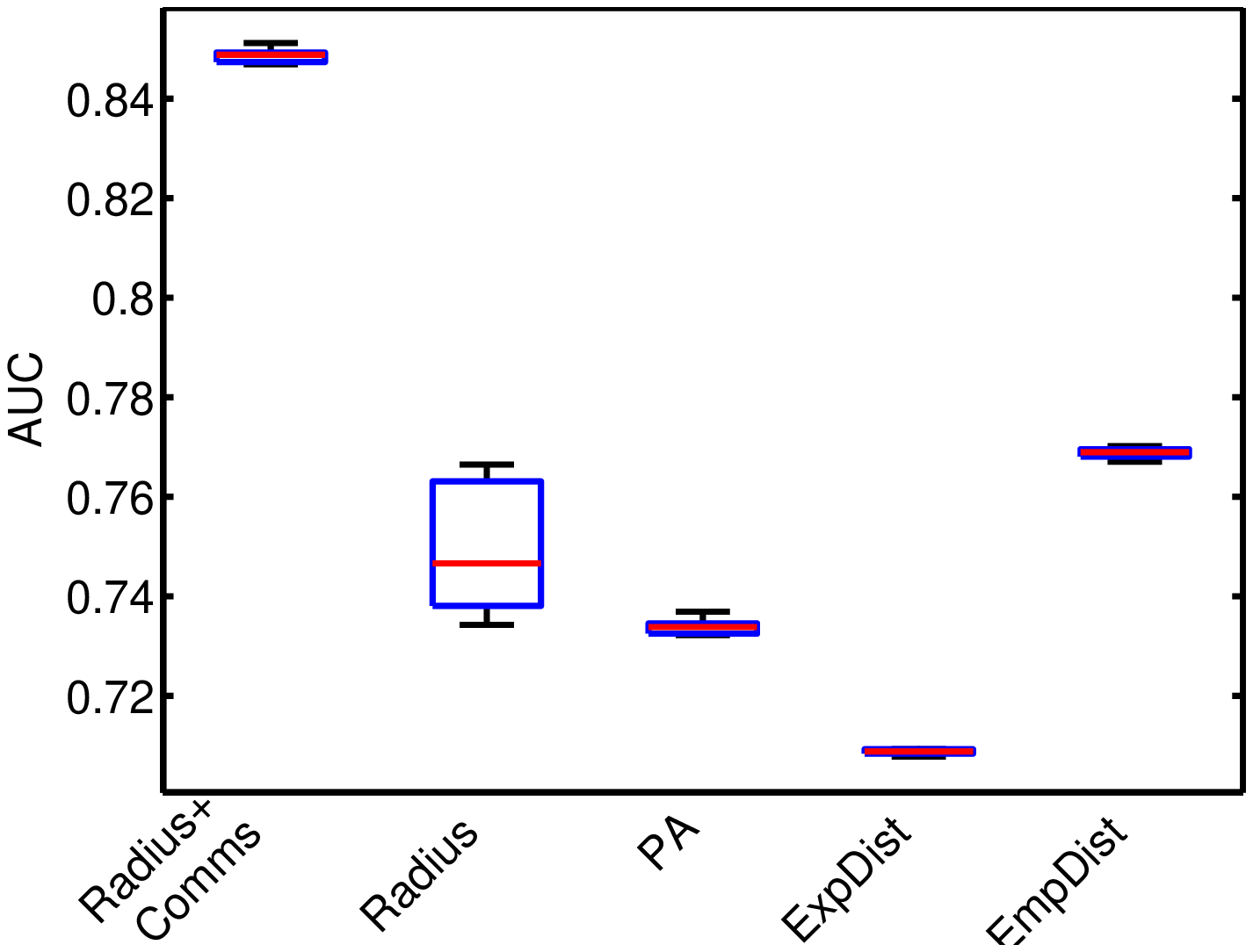}
	}
	\subfigure[Gowalla]{
	    \label{fig:gowalla_auc}
	    \centering
	    \includegraphics[keepaspectratio, width=0.22\textwidth]{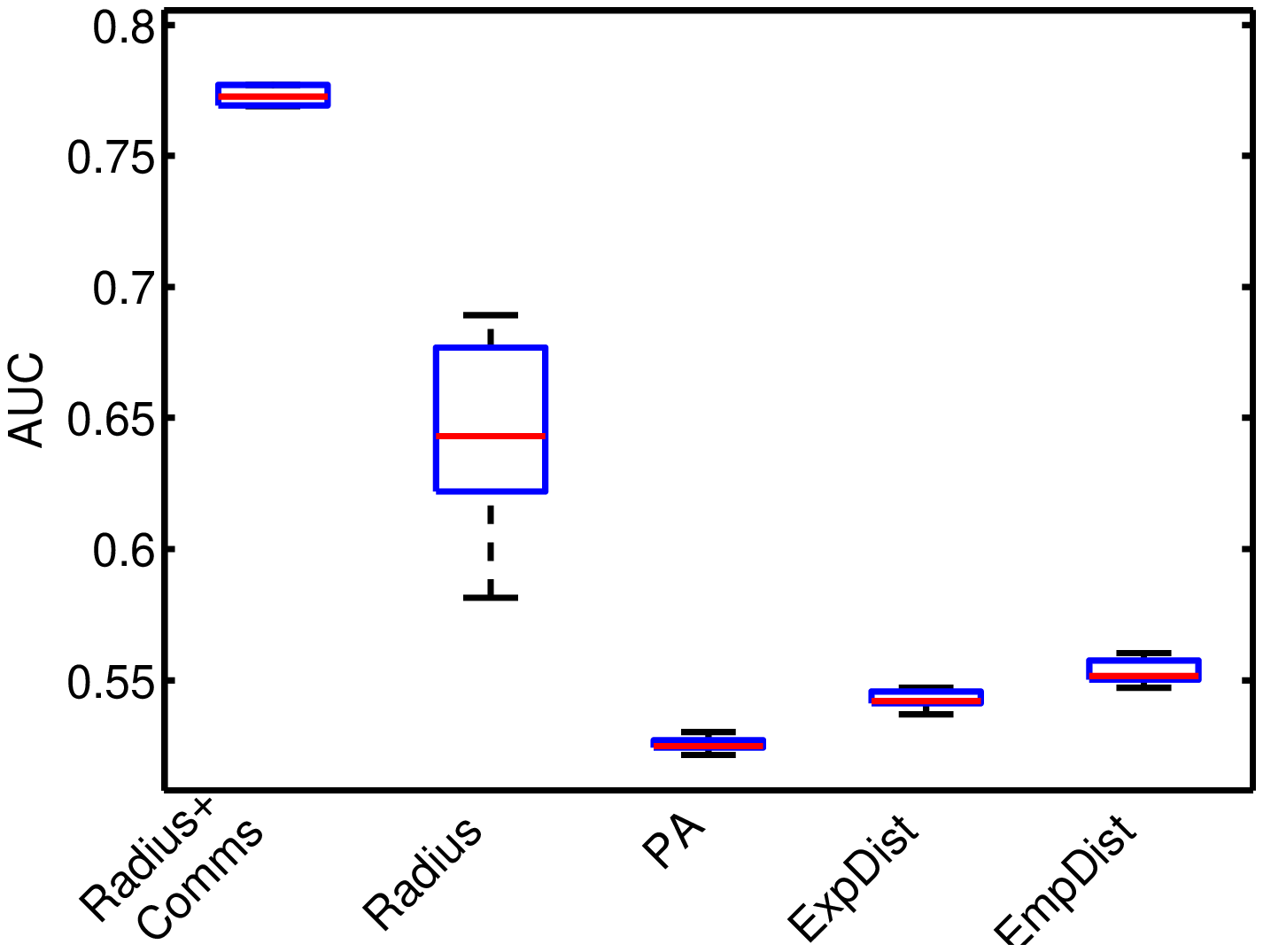}
	}
	\subfigure[CA Internet]{
	    \label{fig:internet_auc}
	    \centering
	    \includegraphics[keepaspectratio, width=0.22\textwidth]{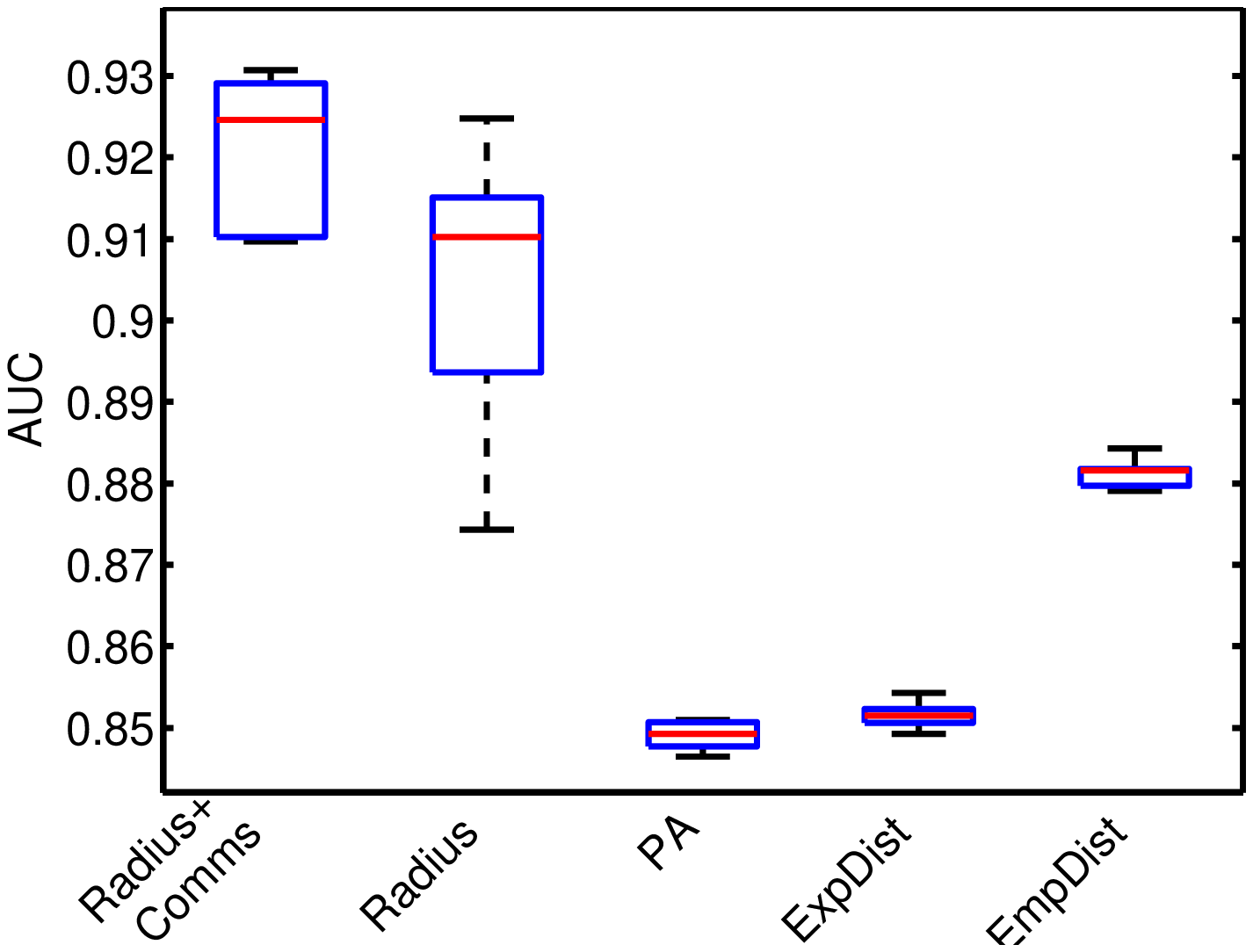}
	}
	\subfigure[US Airline]{
	    \label{fig:airline_auc}
	    \centering
	    \includegraphics[keepaspectratio, width=0.22\textwidth]{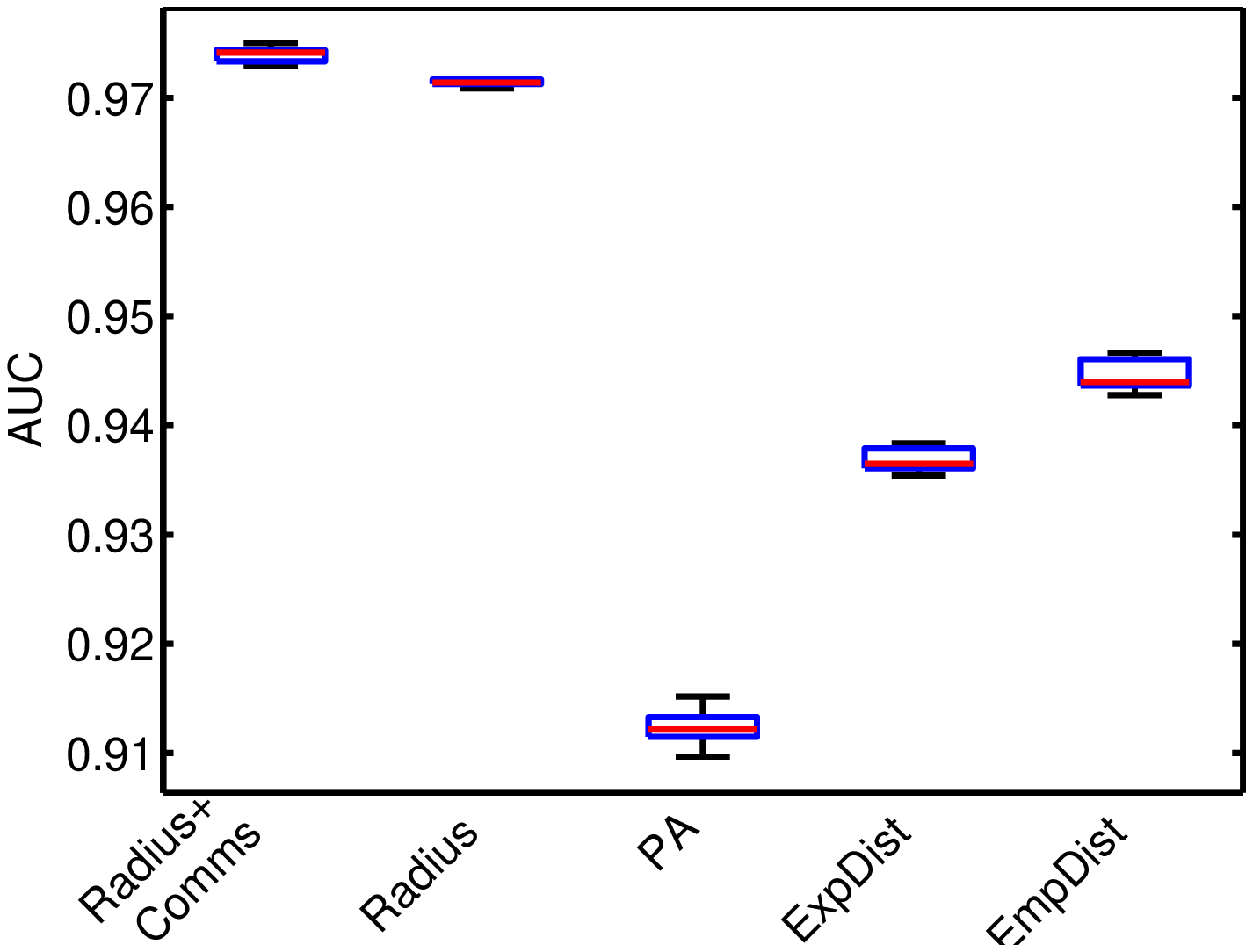}
	}
	\caption{Link prediction AUC over $10$-fold cross validation.}
 \label{fig:auc}
\end{figure*}
To evaluate the link prediction quality of the different methods, we employ area under the receiver operating characteristics (ROC) curve (see~\cite{Fawcett2006} for more details). Figure~\ref{fig:auc} shows the area under the ROC curve (AUC) aggregated over the $10$-folds for each dataset. From these results, we notice several interesting trends. First, the preferential attachment model (PA) (i.e. completely ignoring space) performs surprisingly well, with AUC values typically over $75\%$. Thus, while space certainly plays an important role in the formation of links in these datasets, node popularity is certainly an influential factor in determining network topology which must be taken into consideration. Second, \textit{EmpDist} consistently outperforms both \textit{PA} and \textit{ExpDist}. Additionally, \textit{ExpDist} performs only marginally better than \textit{PA}, except for in the \textit{C. elegans} network where it actually has worse performance. This is likely due to the fact that the true link distance distributions is not actually exponential, as we showed in our earlier analysis.

Lastly, \textit{Radius} typically achieves better predictions than \textit{EmpDist}, though with much higher variability (over the $10$-folds). This is intuitive, since the radii provide more flexibility at the cost of additional model variables which need to be inferred. By accounting for additional community structure within the networks, \textit{Radius+Comms}, provides a substantial improvement over \textit{Radius} in all of the networks. In all of the networks except Internet, we also notice that \textit{Radius+Comms} has much lower variance in its AUC (over the different folds) than \textit{Radius}. This can be attributed to the fact that pairs of nodes between which a link was uncertain in the \textit{Radius} model are likely to be fixed by adding these nodes to the same community, thus explaining part of the link structure more robustly. The high variance in the Internet network is the result of few communities being detected. We investigate the resulting communities in more depth in section~\ref{sec:comms}.

\begin{figure*}[ht]
	\centering
	\subfigure[\textit{C. elegans}]{
	    \label{fig:celegans_auc_dist}
	    \centering
	    \includegraphics[keepaspectratio, width=0.22\textwidth]{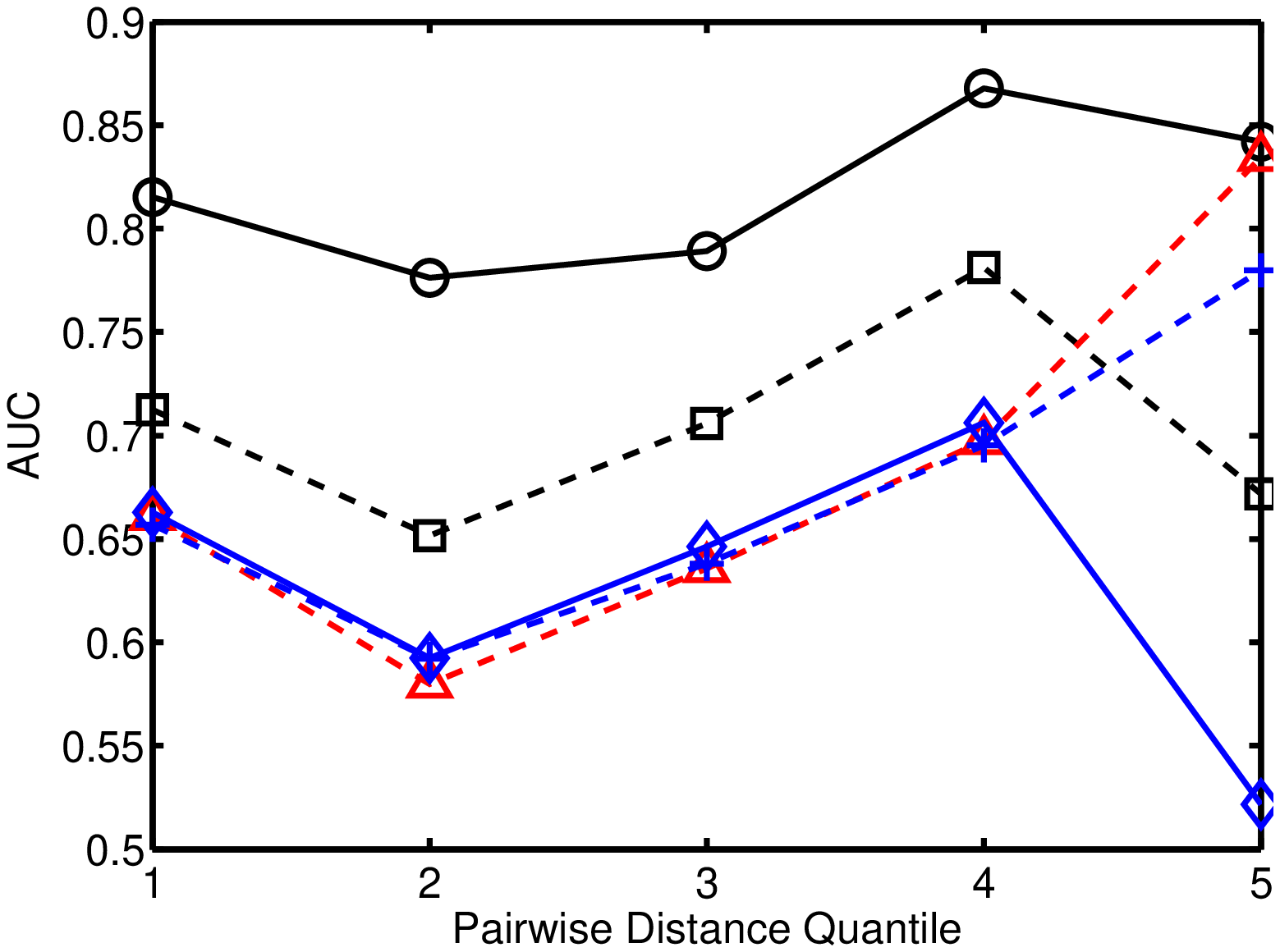}
	}
	\subfigure[Gowalla]{
	    \label{fig:gowalla_auc_dist}
	    \centering
	    \includegraphics[keepaspectratio, width=0.22\textwidth]{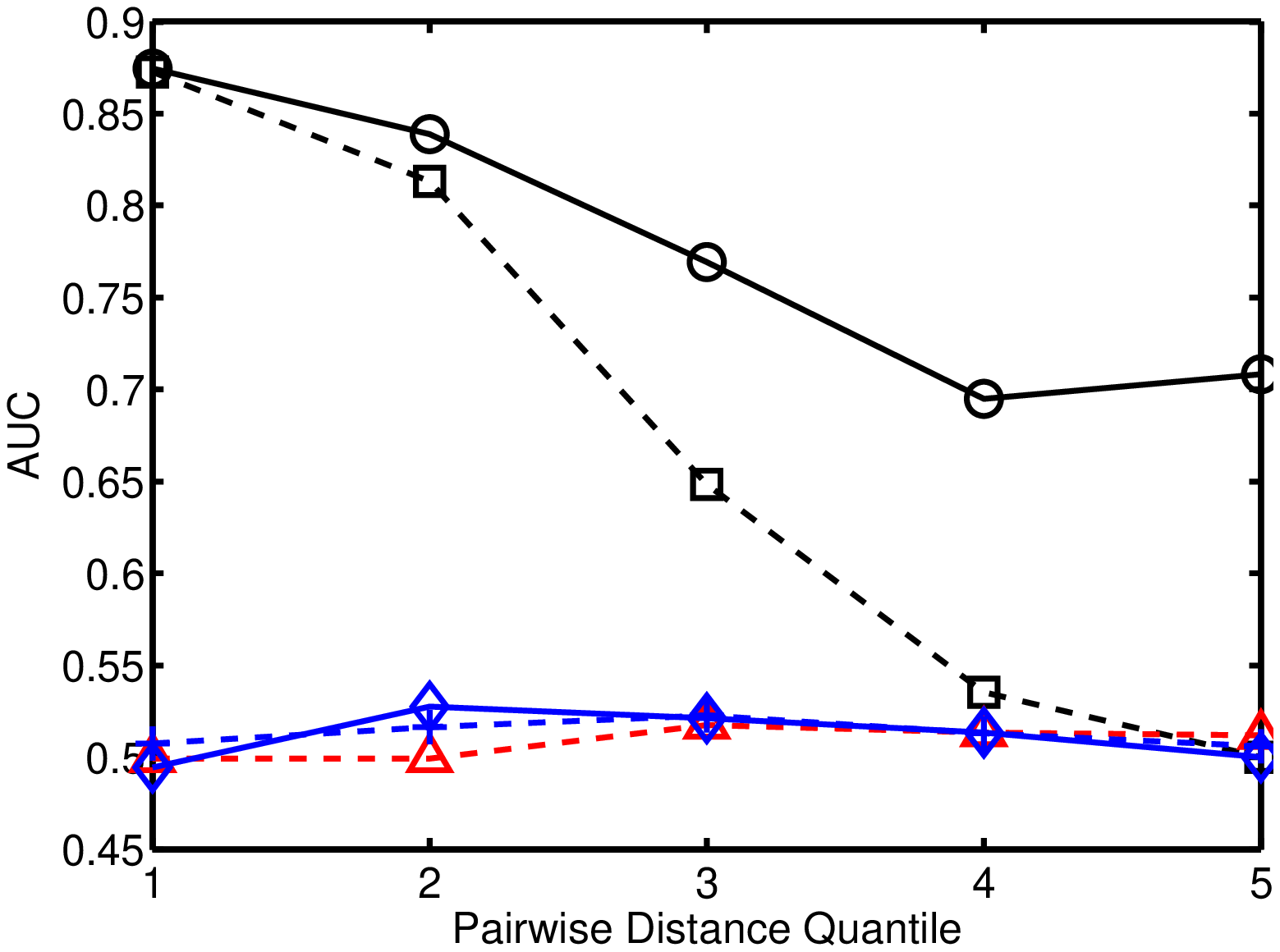}
	}
	\subfigure[CA Internet]{
	    \label{fig:internet_auc_dist}
	    \centering
	    \includegraphics[keepaspectratio, width=0.22\textwidth]{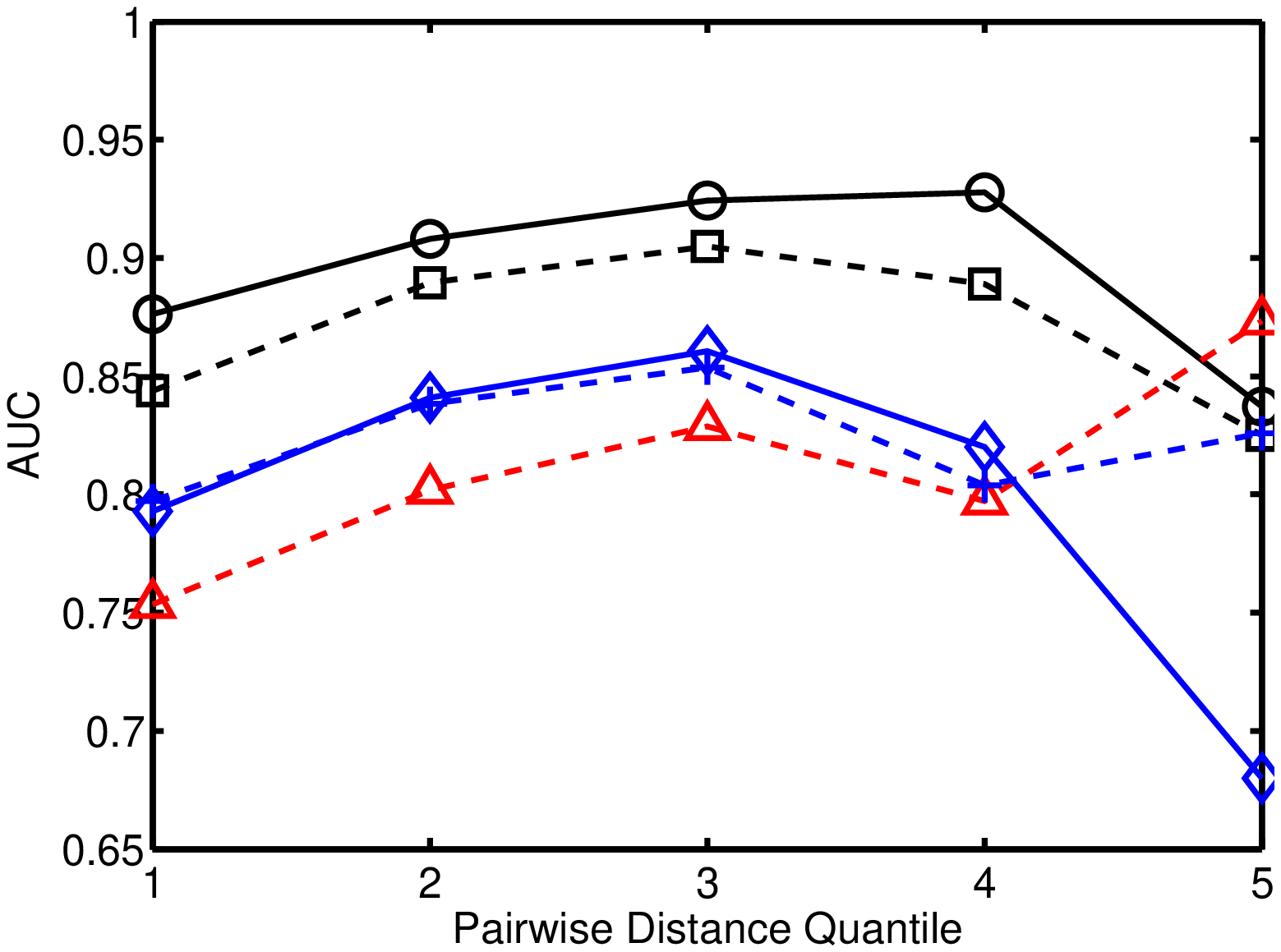}
	}
	\subfigure[US Airline]{
	    \label{fig:airline_auc_dist}
	    \centering
	    \includegraphics[keepaspectratio, width=0.22\textwidth]{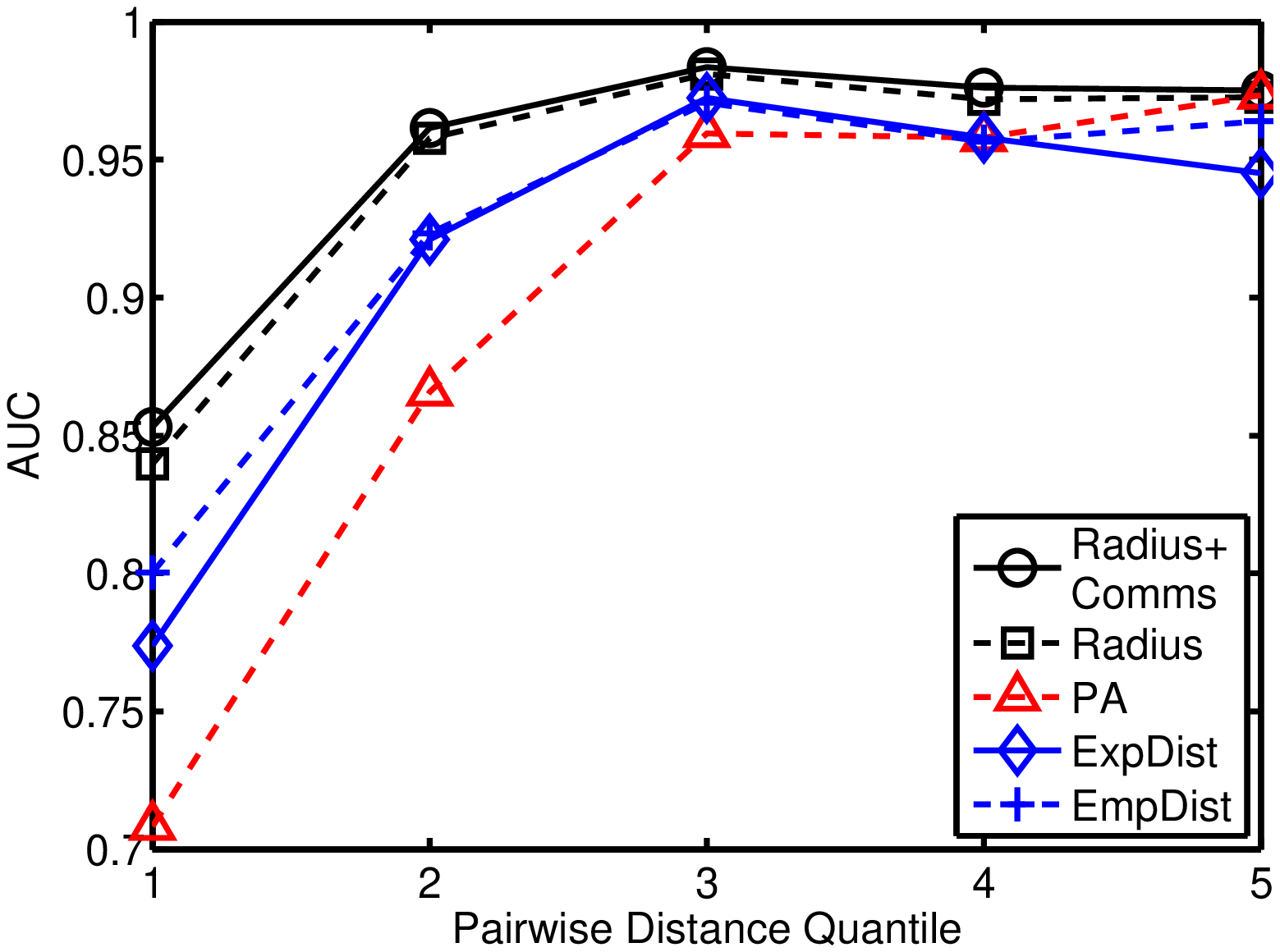}
	}
	\caption{AUC measured over separate quantiles of the test data, split by the pairwise distance between the nodes for which a link is being predicted. The quantiles are shown on the x-axis, where $1$ contains all node-pairs that are close together, and $5$ contains those that are separated by the greatest distances.}
 \label{fig:auc_dist}
\end{figure*}

\begin{figure*}[ht]
	\centering
	\subfigure[\textit{C. elegans}]{
	    \label{fig:celegans_auc_deg}
	    \centering
	    \includegraphics[keepaspectratio, width=0.22\textwidth]{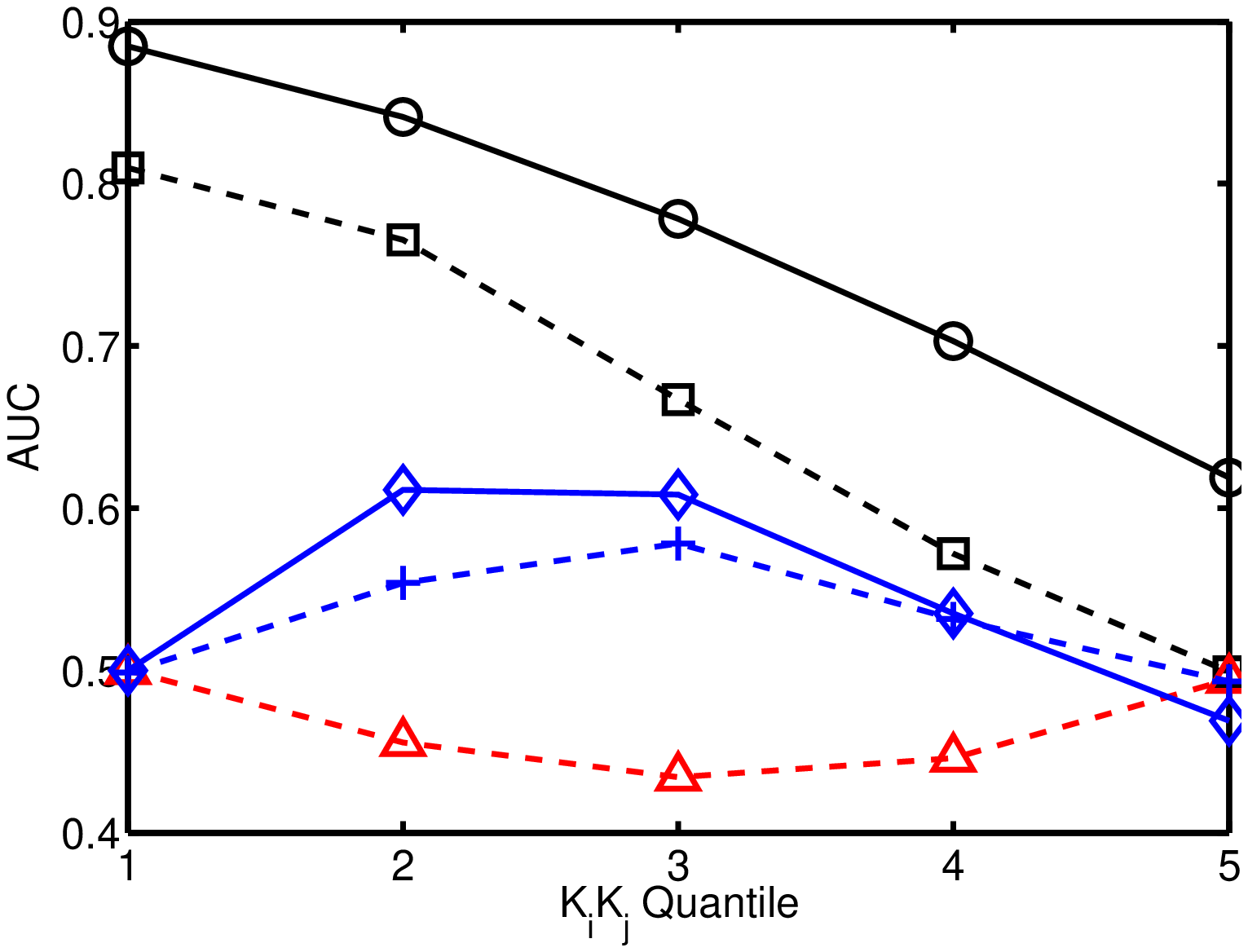}
	}
	\subfigure[Gowalla]{
	    \label{fig:gowalla_auc_deg}
	    \centering
	    \includegraphics[keepaspectratio, width=0.22\textwidth]{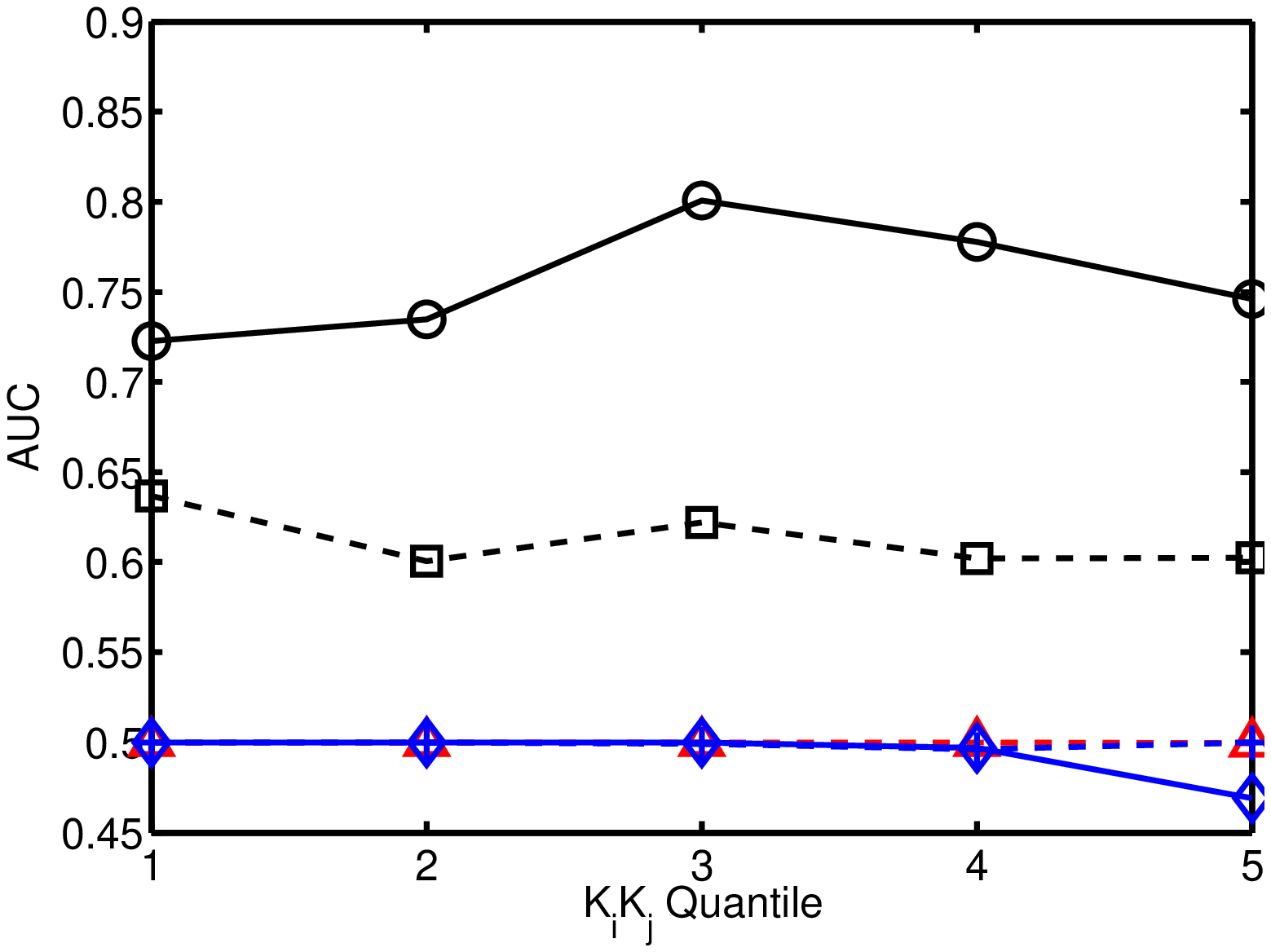}
	}
	\subfigure[CA Internet]{
	    \label{fig:internet_auc_deg}
	    \centering
	    \includegraphics[keepaspectratio, width=0.22\textwidth]{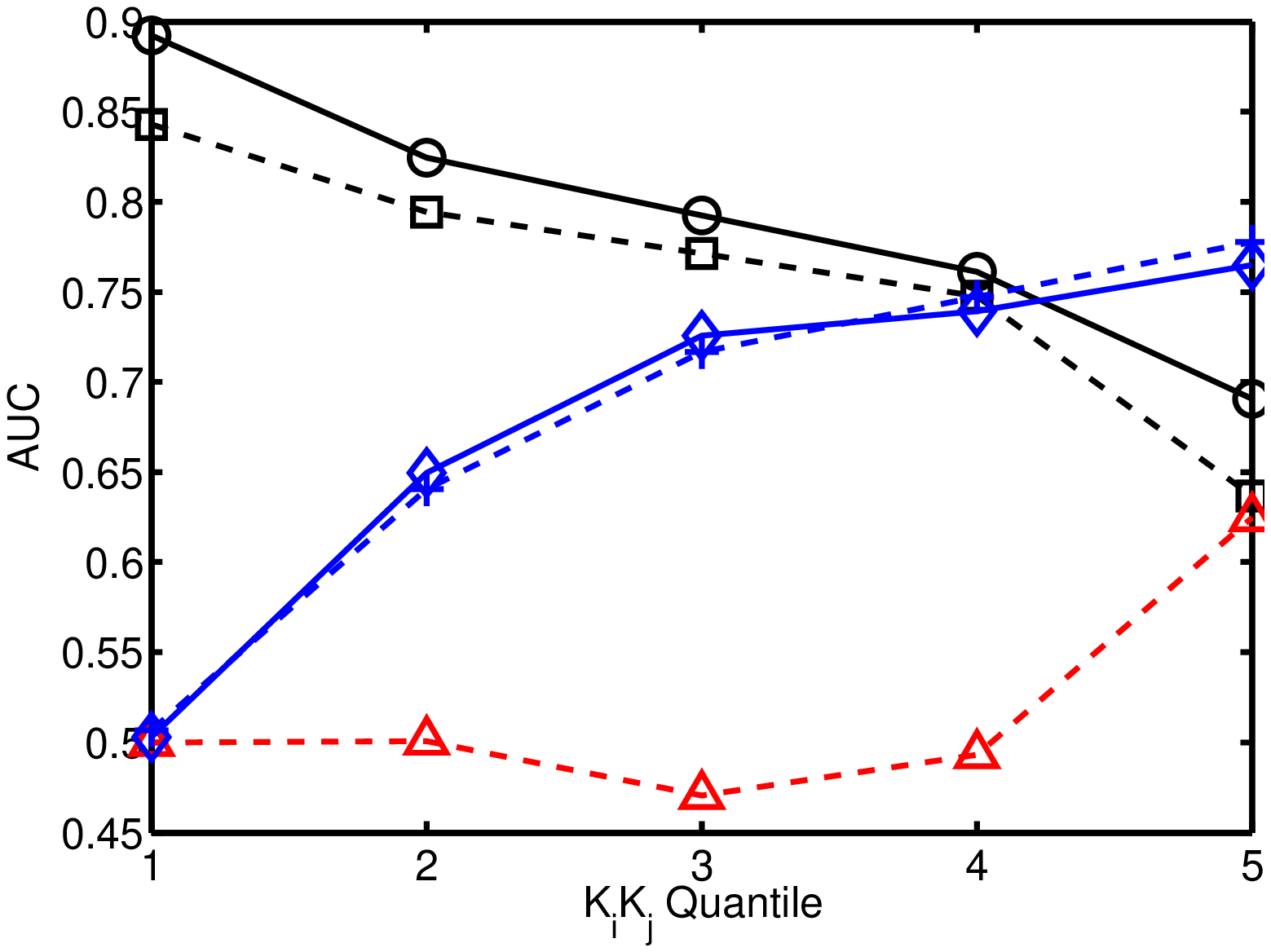}
	}
	\subfigure[US Airline]{
	    \label{fig:airline_auc_deg}
	    \centering
	    \includegraphics[keepaspectratio, width=0.22\textwidth]{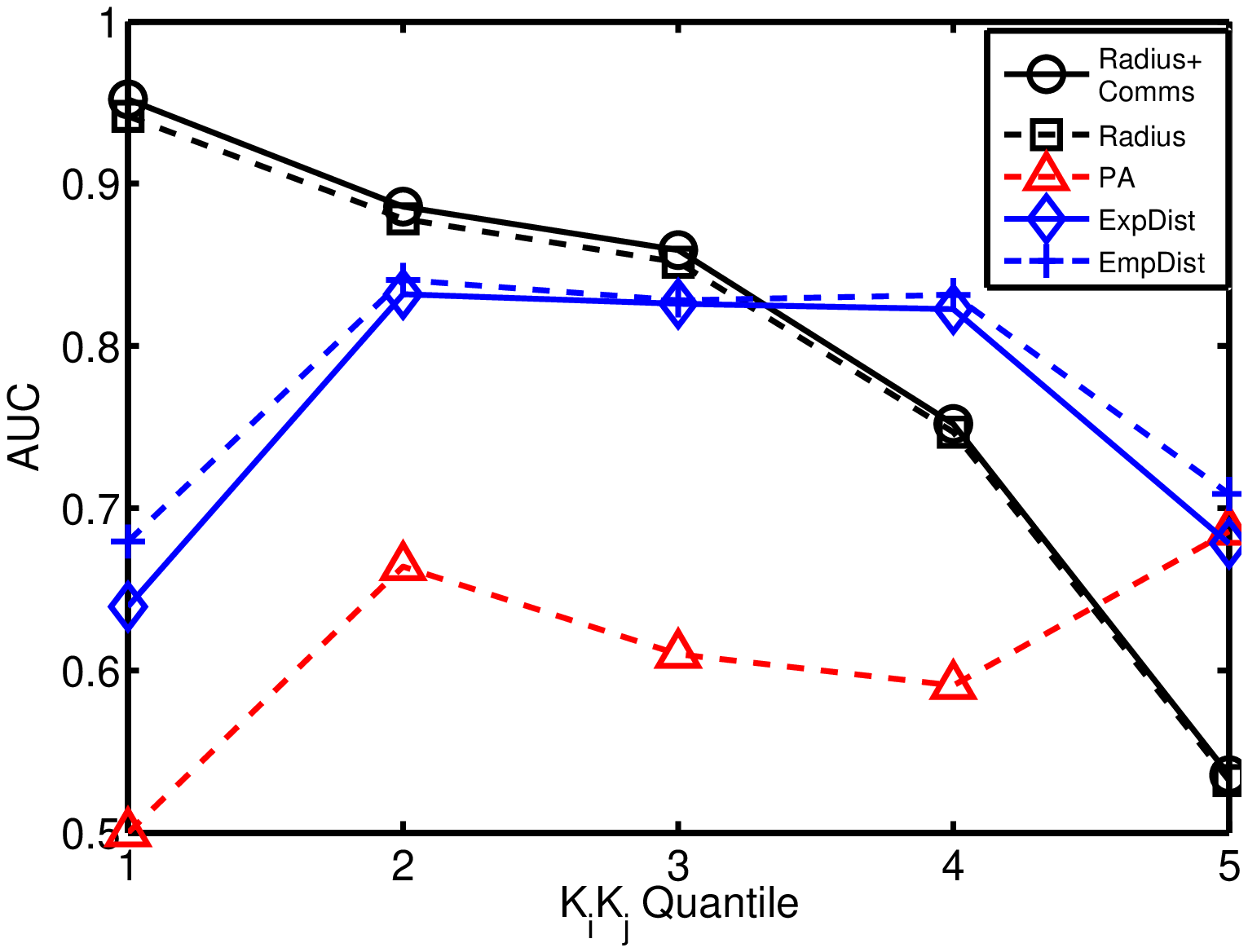}
	}
	\caption{AUC measured over separate quantiles of the test data, split by the combined degrees of the nodes for which a link is being predicted, $k_i k_j$. The quantiles are shown on the x-axis, where $1$ contains all node-pairs in which both nodes have low degree and $5$ contains those in which both nodes have very high degrees.}
 \label{fig:auc_deg}
\end{figure*}

Next, we break down the links according to distance and node degrees to further understand our model's performance. We split the test data into $5$ quantiles based on pairwise node distance and degree, then compute the AUC over each quantile. The quantiles are computed such that there is an even split of links (i.e. true positives) in the testing data into each bin. Figures~\ref{fig:auc_dist} and~\ref{fig:auc_deg} show our results for splits based on pairwise distance and node degree respectively.

Comparing the methods by pairwise distance shows that the \textit{Radius} and \textit{Radius+Comms} models consistently provide higher AUC scores. The only surprise comes from the \textit{C. elegans} and Internet networks at the largest distances, where \textit{Radius} declines while \textit{PA} and \textit{EmpDist} both improve. Because PA improves in this quantile, it suggests that these links may be explained by the node popularity alone. Whereas the \textit{Radius} model is putting too much weight on the distance between these nodes, the other models, with much weaker spatial components, capture these connections due to the popularity of the nodes. The shortcomings in the \textit{Radius} model seem to be overcome in \textit{Radius+Comms}, because the added community variables are able to help explain long distance connections.

Splitting the test data by combined node degrees shows an interesting trend in that the preferential attachment based models are universally bad at predicting edges between nodes with low degrees. This is because the primary source of information used for link prediction in these models is the node degree. Thus if a node is observed as having few connections, it is unlikely to have any more connections. In contrast, the \textit{Radius} model encapsulates information about the network structure local to each node, which is critical to providing accurate predictions for these nodes. For example, if a node is observed to have only one connection but is in a region of low density (i.e. there are few nodes nearby), then any connection made with this node will be further away than the same node in a region of higher density. Whereas the other methods employ a global function of distance which would penalize this node for making such a connection, the radius in our model captures that this is normal given the node's surroundings.

The amount of improvement in link prediction quality our models achieve on low-degree nodes is especially promising. Due to the fact that many nodes are likely to have low degrees (since many networks follow the power-law degree distribution) and network structure alone provides very little information about these nodes, our modeling approach offers a substantial advantage over other techniques. Furthermore, these results emphasize the importance of accurately modeling the link-distance cost function.


\subsection{Community Detection}
\label{sec:comms}
In this section, we investigate the applicability of our models to the task of community detection in spatial networks. We compare the resulting communities identified by our \textit{Radius+Comms} model with previous methods~\cite{Cerina2012, Expert2011}. Additionally, we also use the \textit{Radius} model as a the null comparison within modularity optimization~\cite{Newman2006}. Since no ground truth exists for the community structure in these networks, we provide a pairwise comparison of the different methods. We measure the consistency of the resulting communities across all of the different methods using normalized mutual information (NMI)~\cite{Estevez2009}. By analyzing the similarity of the identified community structures, we show that our proposed model, \textit{Radius+Comms}, captures only the very strongly connected groups of nodes. These are the communities which persist, despite the differences in the clustering objective functions (or the null models).

\begin{table}
	\centering
	\subfigure[\textit{C. elegans}]{{\tiny
	\label{tbl:celegans_nmi}
	\begin{tabular}{| p{0.5cm} | p{0.5cm} | p{0.5cm} | p{0.5cm} | p{0.5cm} | p{0.5cm} |}
		\hline
		& Radius & PA & ExpDist & EmpDist & \blue Radius+ Comm\\ \hline
		Radius & \grey & 0.554 & 0.598 & 0.585 & \blue0.691 \\ \hline
		PA & 0.554 & \grey & 0.629 & 0.699 & \blue0.538 \\ \hline
		ExpDist & 0.598 & 0.629 & \grey & 0.693 & \blue0.533 \\ \hline
		EmpDist & 0.585 & 0.699 & 0.693 & \grey & \blue0.525 \\ \hline
		\red Radius+ Comm & \red0.691 & \red0.538 & \red0.533 & \red0.525 & \grey \\ \hline
	\end{tabular}}}

	\subfigure[Gowalla]{{\tiny
	\label{tbl:gowalla_nmi}
	\begin{tabular}{| p{0.5cm} | p{0.5cm} | p{0.5cm} | p{0.5cm} | p{0.5cm} | p{0.5cm} |}
		\hline
		& Radius & PA & ExpDist & EmpDist &  \blue Radius+ Comm\\ \hline
		Radius & \grey & 0.961 & 0.737 & 0.972 & \blue 0.339 \\ \hline
		PA & 0.927 & \grey & 0.737 & 0.973 & \blue 0.321 \\ \hline
		ExpDist & 0.942 & 0.940 & \grey & 0.740 & \blue 0.436 \\ \hline
		EmpDist & 0.941 & 0.950 & 0.945 & \grey & \blue 0.321 \\ \hline
		\red Radius+ Comm & \red0.961 & \red0.935 & \red0.927 & \red0.934 & \grey  \\ \hline
	\end{tabular}}}

	\subfigure[CA Internet]{{\tiny
	\label{tbl:internet_nmi}
	\begin{tabular}{| p{0.5cm} | p{0.5cm} | p{0.5cm} | p{0.5cm} | p{0.5cm} | p{0.5cm} |}
		\hline
		& Radius & PA & ExpDist & EmpDist & \blue Radius+ Comm\\ \hline
		Radius & \grey & 0.453 & 0.570 & 0.556 & \blue0.089 \\ \hline
		PA & 0.791 & \grey & 0.444 & 0.482 & \blue0.099 \\ \hline
		ExpDist & 0.856 & 0.786 & \grey & 0.531 & \blue0.092 \\ \hline
		EmpDist & 0.846 & 0.803 & 0.884 & \grey & \blue0.088 \\ \hline
		\red Radius+ Comm & \red0.881 & \red0.802 & \red0.870 & \red0.880 & \grey \\ \hline
	\end{tabular}}}

	\subfigure[US Airline]{{\tiny
	\label{tbl:airline_nmi}
	\begin{tabular}{| p{0.5cm} | p{0.5cm} | p{0.5cm} | p{0.5cm} | p{0.5cm} | p{0.5cm} |}
		\hline
		& Radius & PA & ExpDist & EmpDist & \blue Radius+ Comm\\ \hline
		Radius & \grey & 0.542 & 0.609 & 0.646 & \blue0.140 \\ \hline
		PA & 0.712 & \grey & 0.493 & 0.542 & \blue0.132 \\ \hline
		ExpDist & 0.790 & 0.663 & \grey & 0.639 & \blue0.150 \\ \hline
		EmpDist & 0.829 & 0.750 & 0.897 & \grey & \blue0.144 \\ \hline
		\red Radius+ Comm & \red0.882 & \red0.751 & \red0.852 & \red0.885 & \grey \\ \hline
	\end{tabular}}}

	\caption{Agreement between community detection methods. The top triangular matrix contains normalized mutual information (NMI) scores comparing the resulting communities between the different methods. The bottom triangular matrix shows NMI over just the subset of nodes that \textit{Radius+Comms} placed into a community. The number of nodes considered for each network were: \subref{tbl:celegans_nmi} 277, \subref{tbl:gowalla_nmi} 134, \subref{tbl:internet_nmi} 28, \subref{tbl:airline_nmi} 36. The first four rows (columns) are computed by using the referenced model as the null model and applying modularity optimization~\cite{Newman2006}. The last row (column), with the blue tinted background, is the result of our \textit{Radius+Comms} model, in which the community structure is identified within the model itself.}
  	\label{tbl:comm_agreement}
\end{table}

\begin{figure*}[ht]
    \centering
    \subfigure[PA]{
	\label{fig:ng}
	\centering
	\includegraphics[keepaspectratio, width=0.4\textwidth]{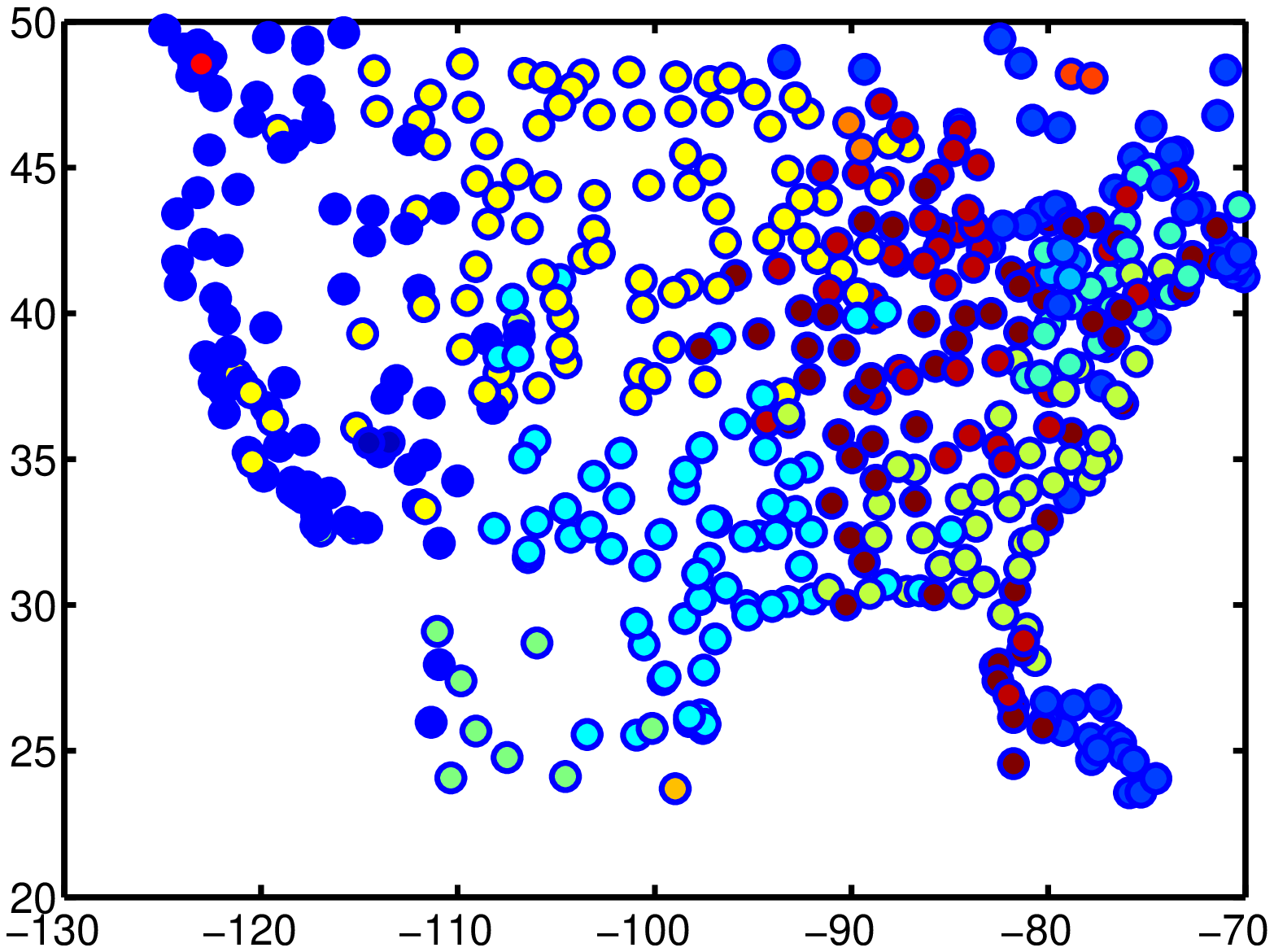}
    }
    \subfigure[ExpDist]{
	\label{fig:arvix}
	\centering
	\includegraphics[keepaspectratio, width=0.4\textwidth]{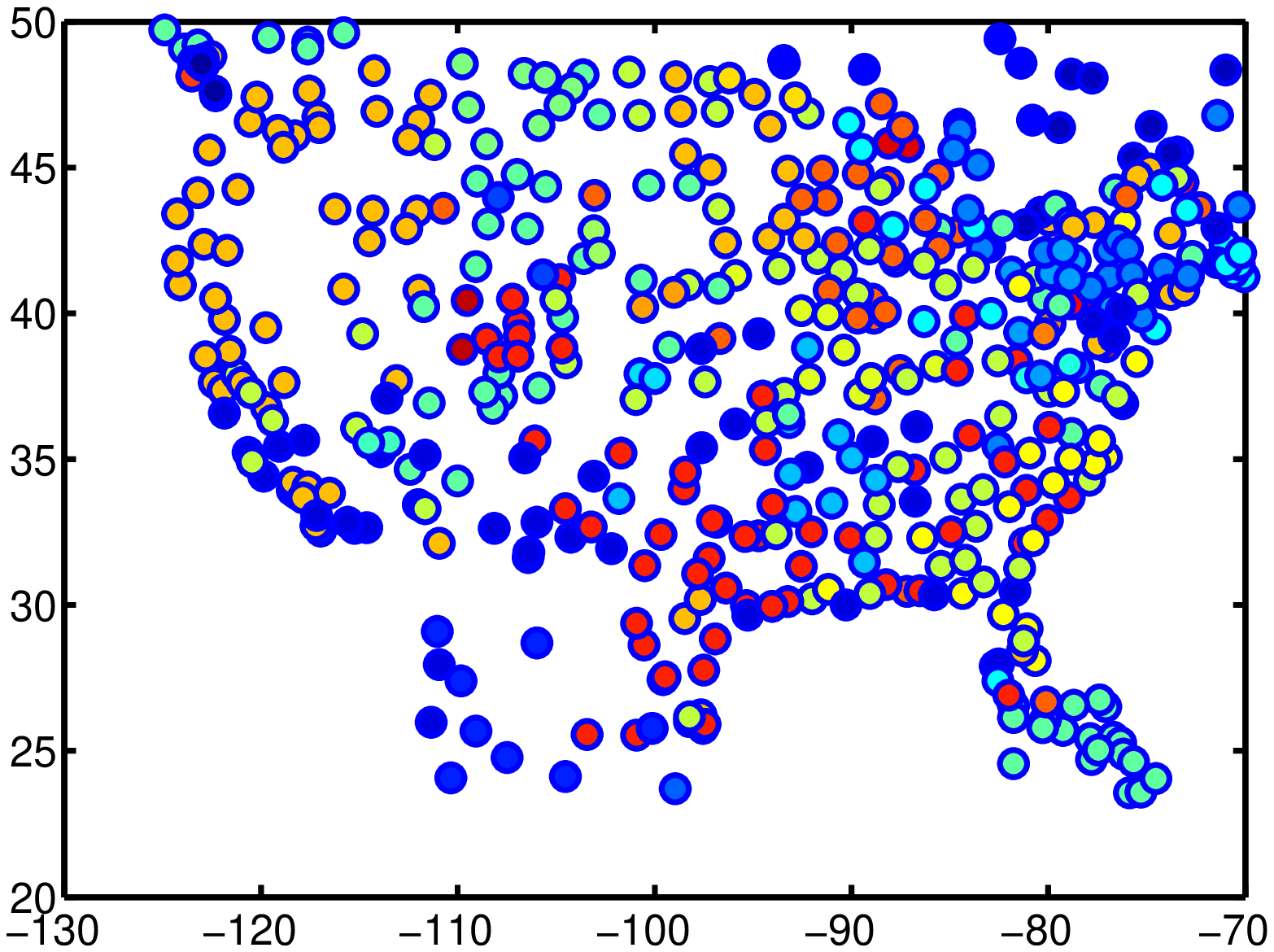}
    }
    \subfigure[EmpDist]{
	\label{fig:pnas}
	\centering
	\includegraphics[keepaspectratio, width=0.4\textwidth]{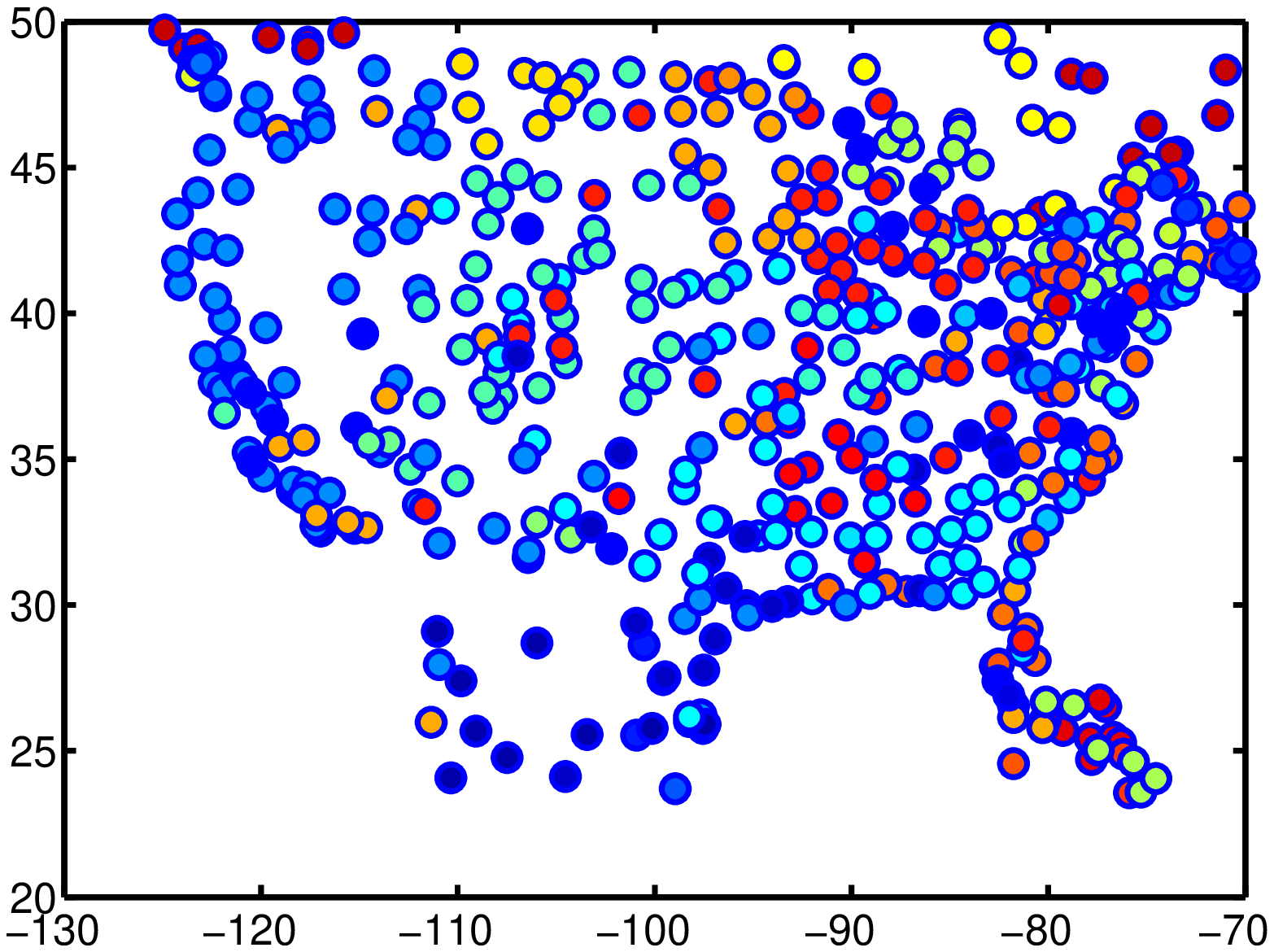}
    }
    \subfigure[Radius+Comms]{
	\label{fig:me}
	\centering
	\includegraphics[keepaspectratio, width=0.4\textwidth]{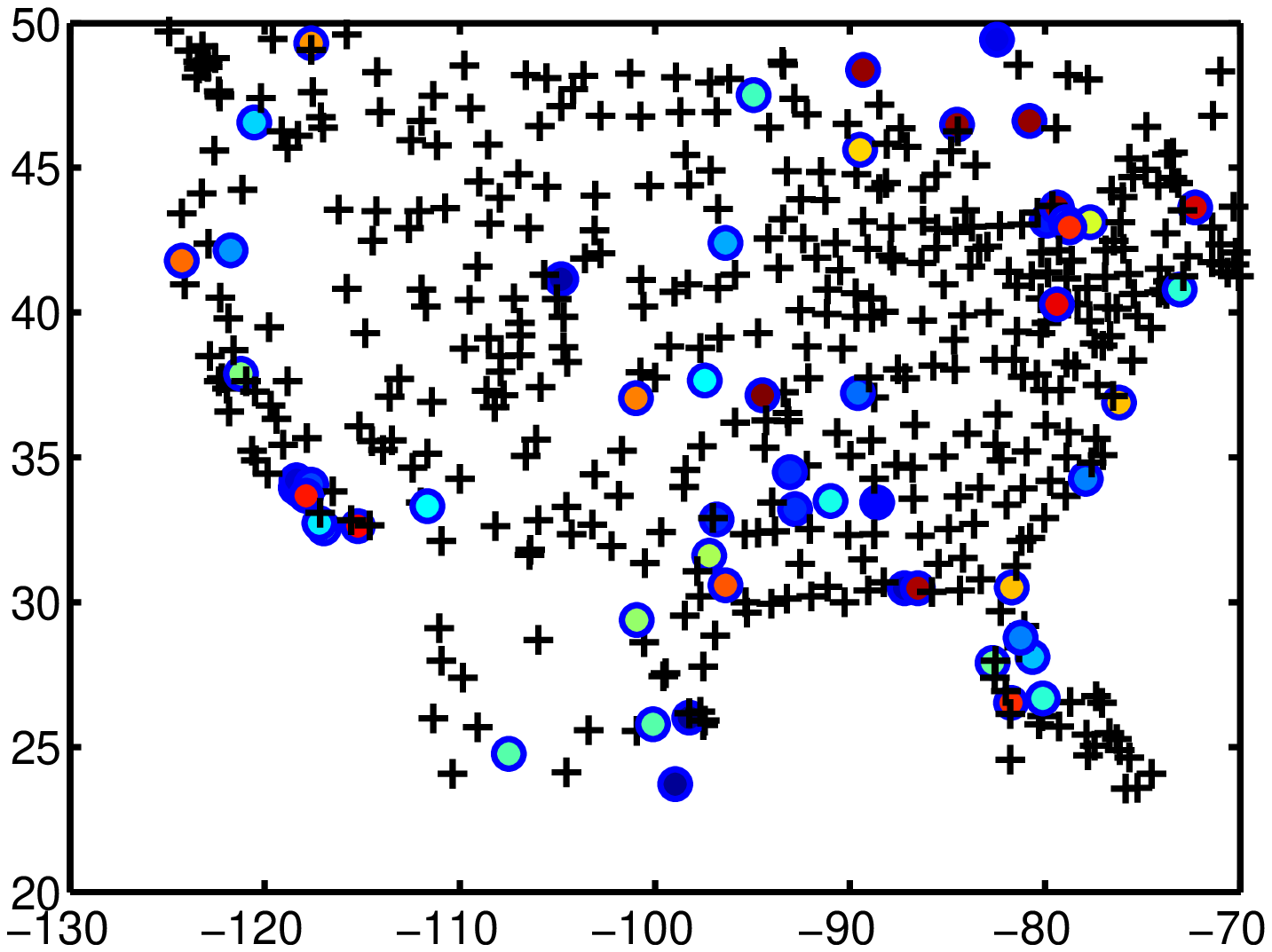}
    }
    \caption{The communities detected by the different methods in the Airline network (best viewed in color). The communities identified by PA show a strong spatial structure, which is mostly maintained in \textit{ExpDist} and \textit{EmpDist} as well, although nodes on the fringe may switch to neighboring communities. In contrast, \textit{Radius+Comms} identifies much fewer, though much more strongly integrated communities (nodes not belonging to any community are shown as black $+$'s) for which it is difficult to identify any real spatial structure.}
    \label{fig:comms}
\end{figure*}

We observe that all of the spatial, modularity-based models tend to produce results more similar to each other than to the basic PA null model. This is intuitive, as each of these models is considering the same additional information about network structure, though they are incorporating this information differently. Additionally, the two baseline spatial null models, \textit{ExpDist} and \textit{EmpDist}, show similar levels of agreement amongst themselves indicating that even relatively small changes in the null model can force nodes on the fringe of a community to switch to another group. This is shown visually in figure~\ref{fig:comms}.

In general, we see very little agreement between the communities discovered using the modularity-based approaches and \textit{Radius+Comms}. This is due to two major differences in the objective function. First, modularity only optimizes within cluster edges and does not explicitly penalize strong connections between clusters. This is in contrast to our method which equally rewards within cluster links as well as penalizes between cluster links. Second, modularity forces all nodes to be placed into a cluster, whereas \textit{Radius+Comms} contains a special \textit{don't care} group for which nodes are unaffected by community structure. This provides additional modeling flexibility in that we can both find instances where community structure helps explain link structure as well as instances where nodes do not appear to be affected (i.e. link structure can be explained by spatial and preferential attachment effects).

However, examining the subset of nodes which are explicitly placed into communities in \textit{Radius+Comms}, we find very strong agreement across all of the clustering methods (bottom half of tables in~\ref{tbl:comm_agreement}). The fact that much of the community structure found using our method persists even when the clustering objective function is modified, indicates that \textit{Radius+Comms} is identifying only the most significant communities. In fact, the importance of the identified community structure is orated by our link prediction results as well. \textit{Radius+Comms} offers substantial improvements over \textit{Radius} in our ability to explain the network structure, and thus predict missing links across all of the data sets.

Upon further inspection, we see that the communities identified by \textit{Radius+Comms} are in fact spatial anomalies. One such example of this is in the Airline network where we find that the Lake Charles Regional Airport in Lake Charles, Louisiana and the Chris Hadfield Airport in Sarnia, Ontario which are placed into the same community. These two airports are separated by more than $1,700$ km, and the airports have a total of $2$ and $1$ recorded connections respectively. Given the size of these airports and the large distance separating them, such a connection is truly not expected.

\begin{figure}[ht]
    \centering
    \subfigure[]{
	\label{fig:ex_comm1}
	\centering
	\includegraphics[keepaspectratio, width=0.22\textwidth]{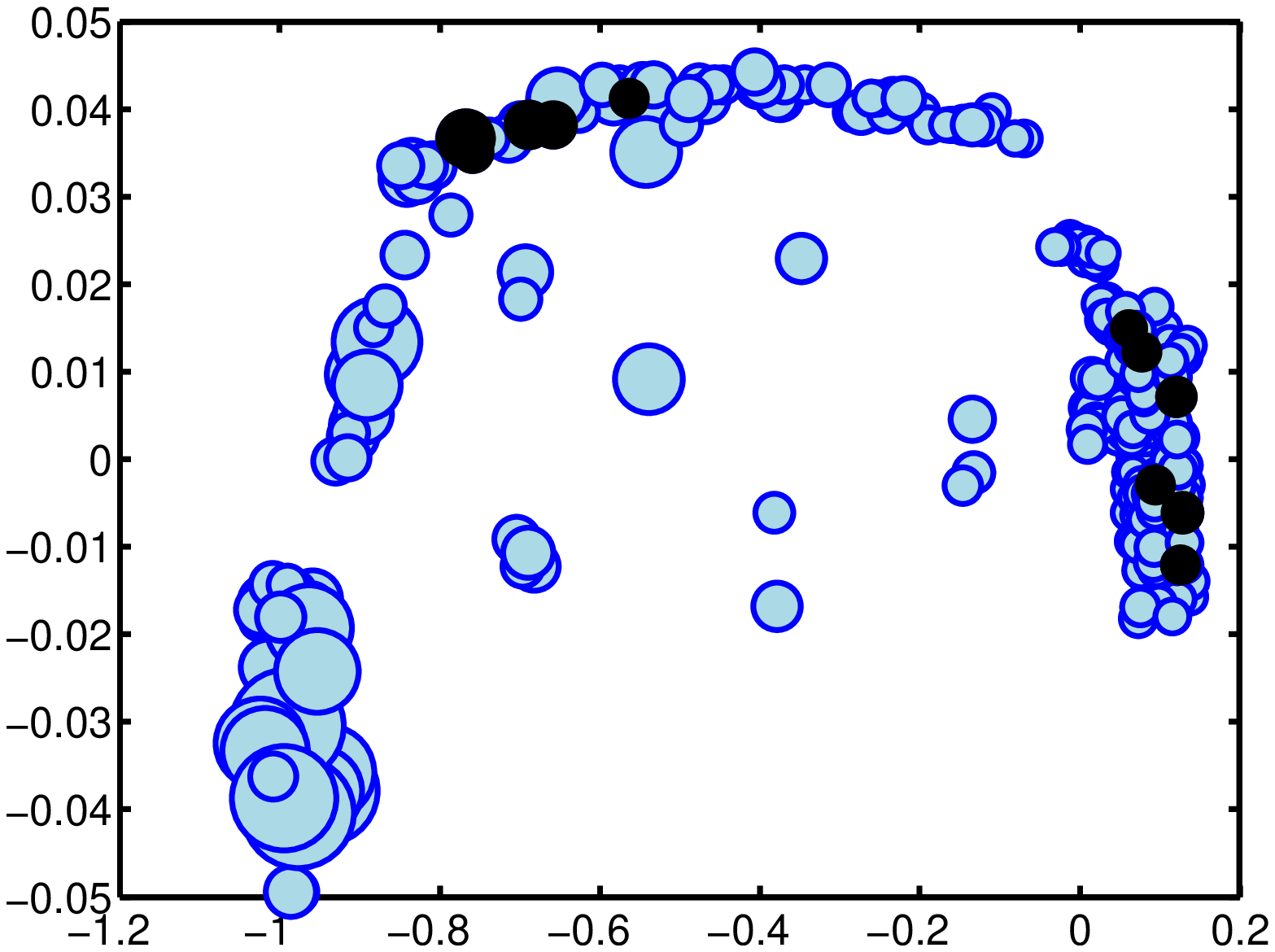}
    }
    \subfigure[]{
	\label{fig:ex_comm2}
	\centering
	\includegraphics[keepaspectratio, width=0.22\textwidth]{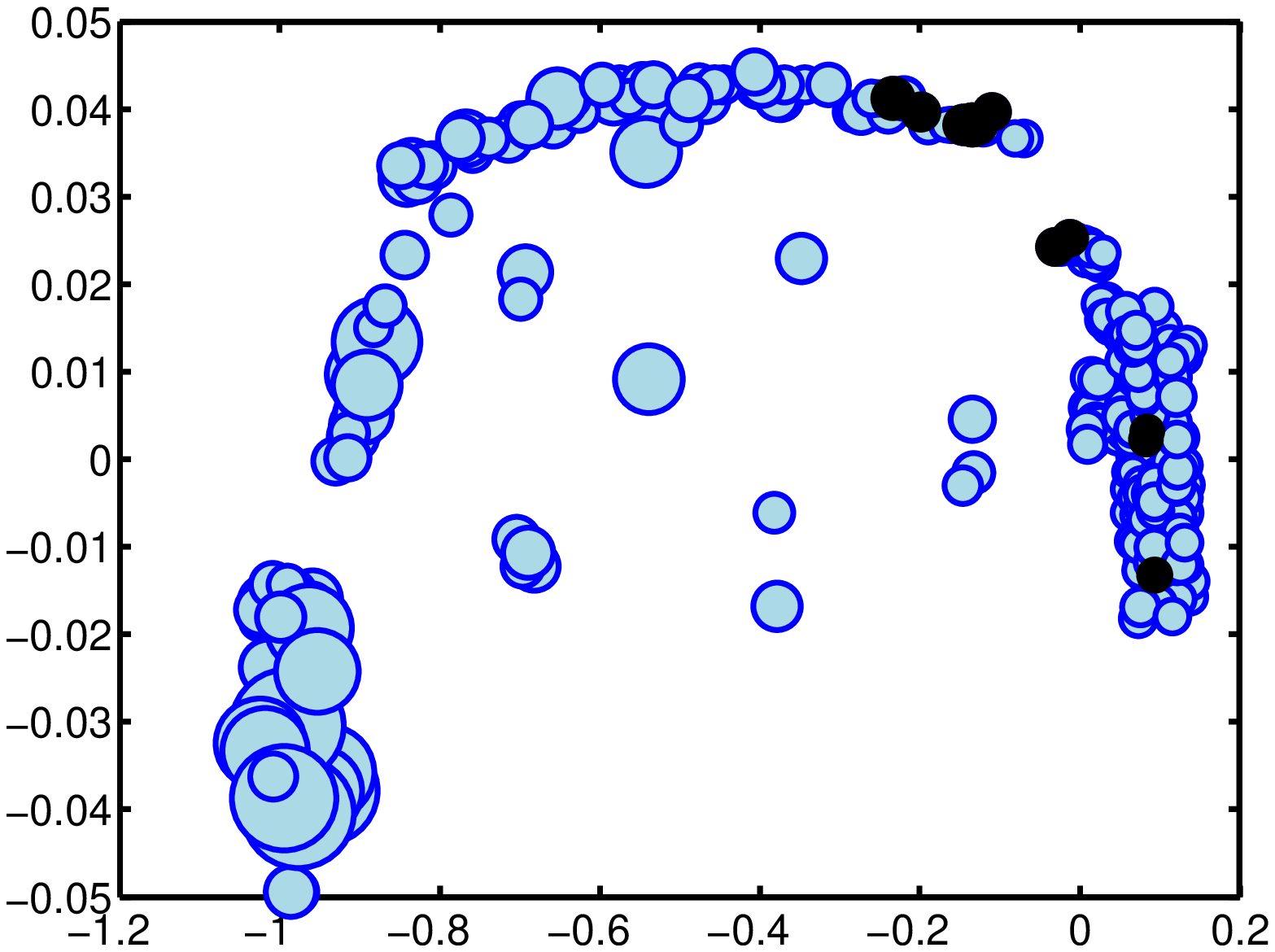}
    }
    \caption{Sample communities, shown as black nodes, identified by \textit{Radius+Comms}.}
    \label{fig:ex_comms}
\end{figure}

Similarly, figure~\ref{fig:ex_comms} shows two example communities identified in the \textit{C. elegans} network. Despite being spatially diverse, both communities are composed of functionally similar neurons. The community in figure~\ref{fig:ex_comm1} includes Ventral cord motor neurons and interneurons which play a role in locomotion. Similarly, the community shown in figure~\ref{fig:ex_comm2} is composed of a mix of mechanosensory and additional ventral cord motor neurons. The functions of these neurons all surround the task of locomotion as well as collision detection~\cite{Donald1997, Varshney2011}. These examples indicate that there is indeed a reasonable level of coherence within the communities.

\subsection{MCMC Analysis}
\label{sec:mcmc}
Lastly, we discuss the convergence and mixing properties of our MCMC algorithm. To guarantee good mixing and quick convergence, we wish to provide a good initialization of the parameters. For each network, we run a short Markov chain and use the maximum a-posterior (MAP) configuration from that run to initialize the model parameters. While we find that we are able to converge quickly for most of the datasets, convergence on the airline network was particularly slow. We observe a large initial jump in the log posterior after the first few iterations when we move from the randomly initialized parameter values into a more coherent configuration.

\begin{figure}[ht]
    \centering
    \subfigure[\textit{C. elegans}]{
	\label{fig:celegans_logpost}
	\centering
	\includegraphics[keepaspectratio, width=0.22\textwidth]{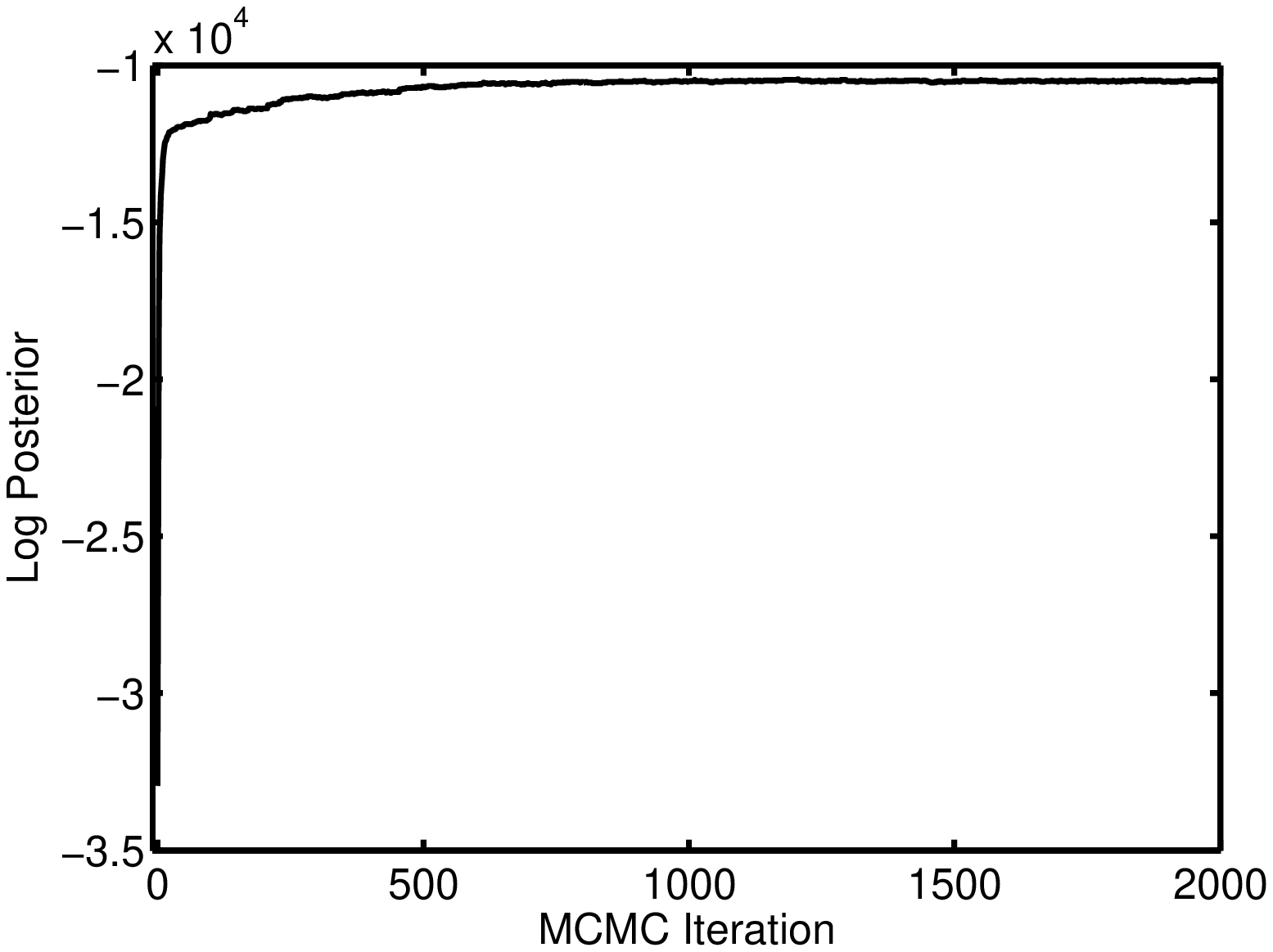}
    }
    \subfigure[Airline]{
	\label{fig:airline_logpost}
	\centering
	\includegraphics[keepaspectratio, width=0.22\textwidth]{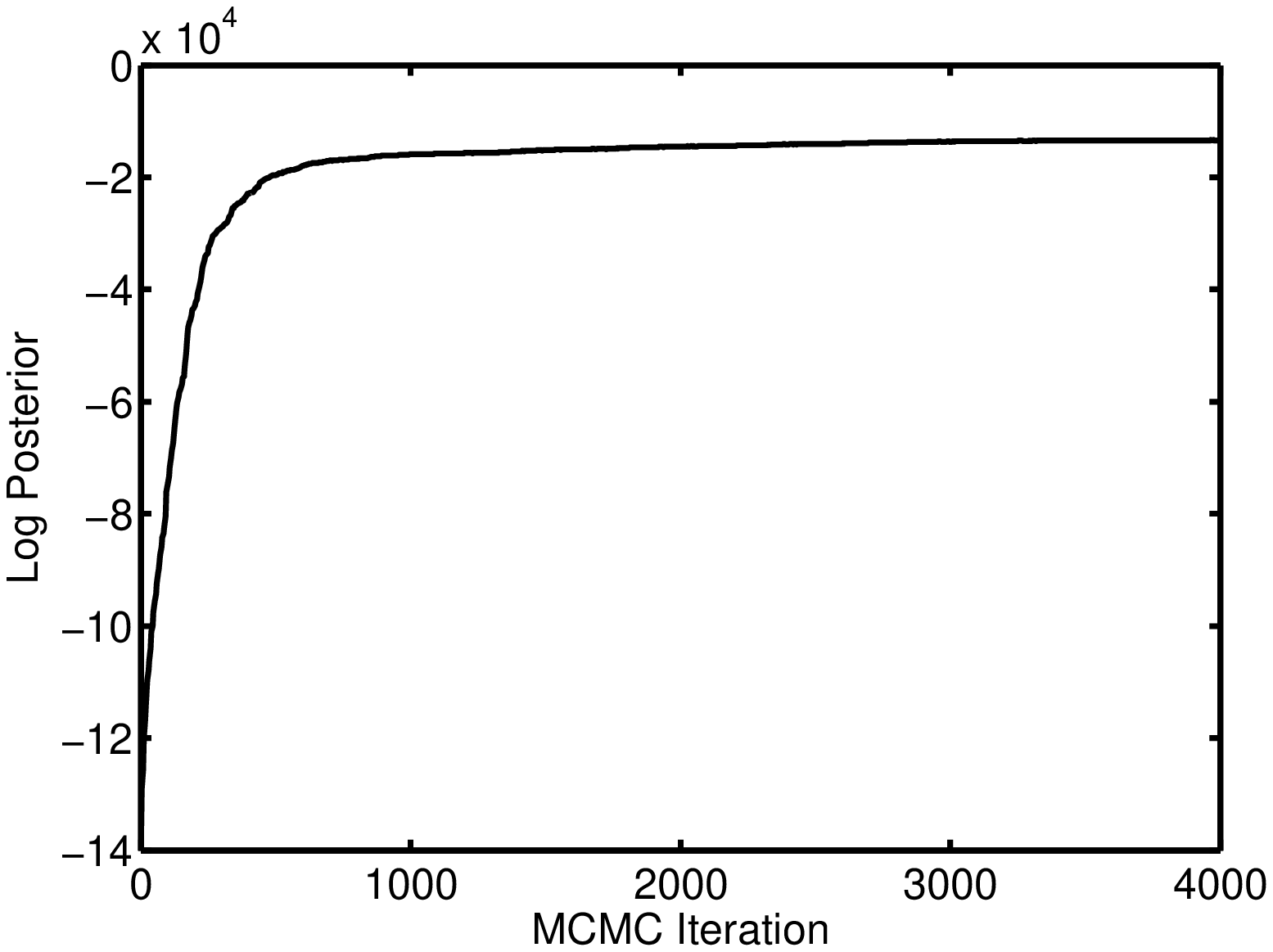}
    }
    \caption{Log posterior trace plots from initialization run of the~\subref{fig:celegans_logpost} \textit{C. elegans} network and the~\subref{fig:airline_logpost} US Airline network. For \textit{C. elegans}, we observe fast convergence of the log posterior in under $2,000$ iterations, whereas for the Airline network, we observe the posterior is still rising, at a very slow rate, past $4,000$ iterations.}
    \label{fig:logpost}
\end{figure}

However, unlike the other networks in which the log-posterior flattens out indicating that we have reached the mode of the distribution, the airline network slowly improves over several thousand iterations until it finally converges into a posterior mode. Such a slow convergence indicates that the posterior distribution may be rather diffuse for the given data and thus several parameter configurations may provide similarly adequate fits for the network. Figure~\ref{fig:logpost} shows the log posterior from the \textit{C. elegans} and US Airline networks. Despite the slow convergence on the Airline network, we still see consistent results across multiple runs.

\begin{figure}[ht]
	\centering
	\subfigure[$\alpha$]{
	    \label{fig:prior_alpha}
	    \centering
	    \includegraphics[keepaspectratio, width=0.30\textwidth]{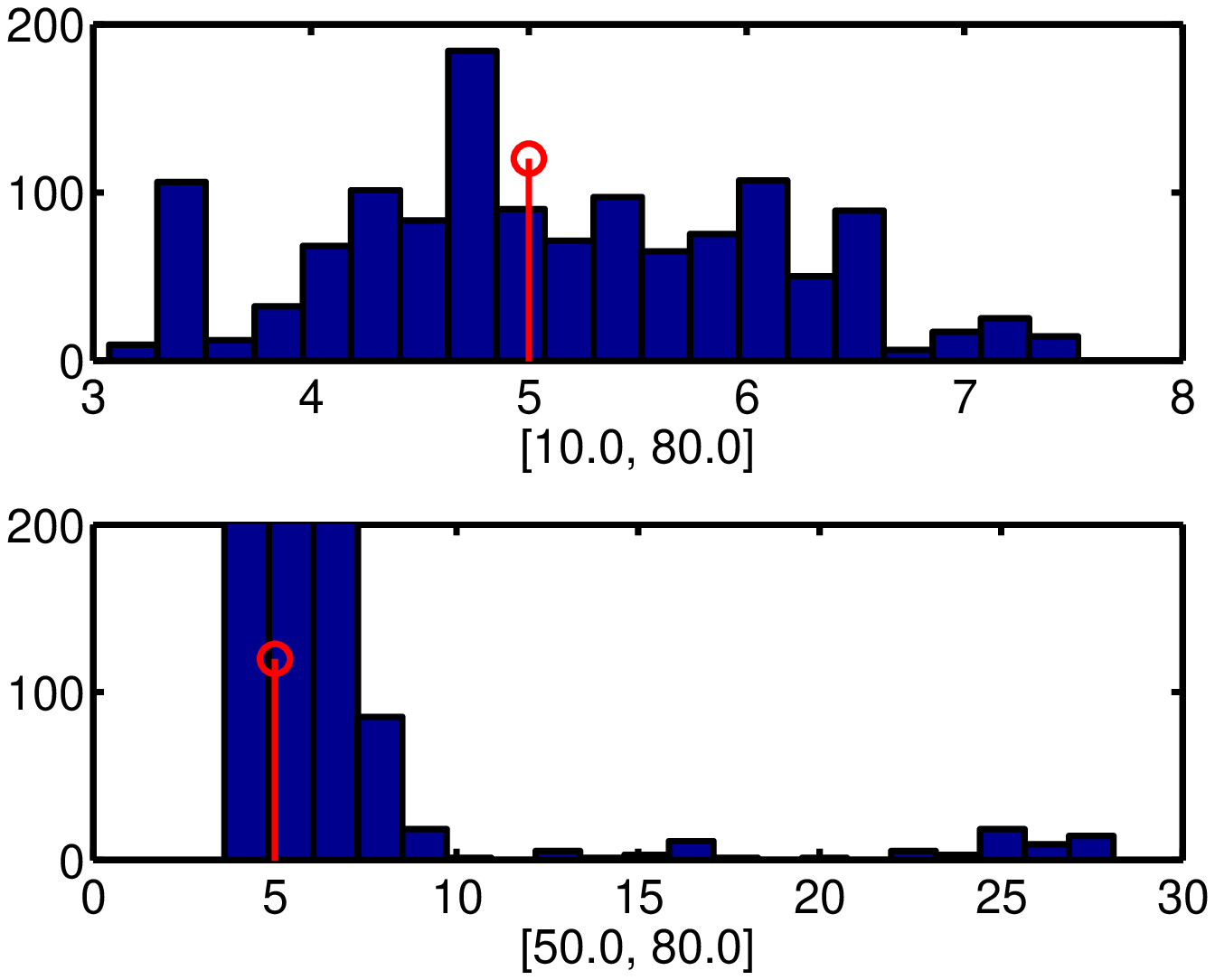}
	}
	\subfigure[$\beta$]{
	    \label{fig:prior_beta}
	    \centering
	    \includegraphics[keepaspectratio, width=0.30\textwidth]{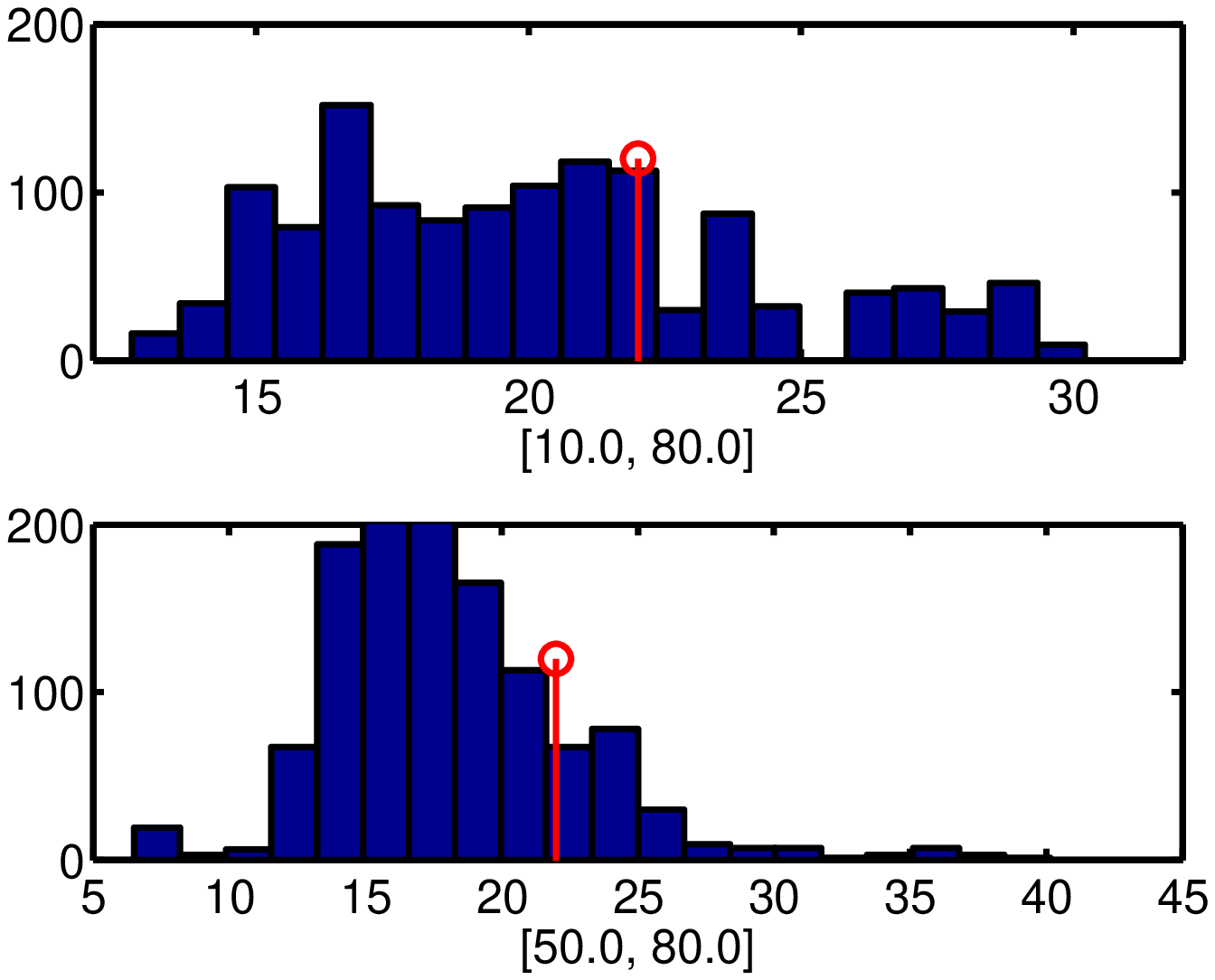}
	}
	\subfigure[Radius]{
	    \label{fig:prior_radius}
	    \centering
	    \includegraphics[keepaspectratio, width=0.30\textwidth]{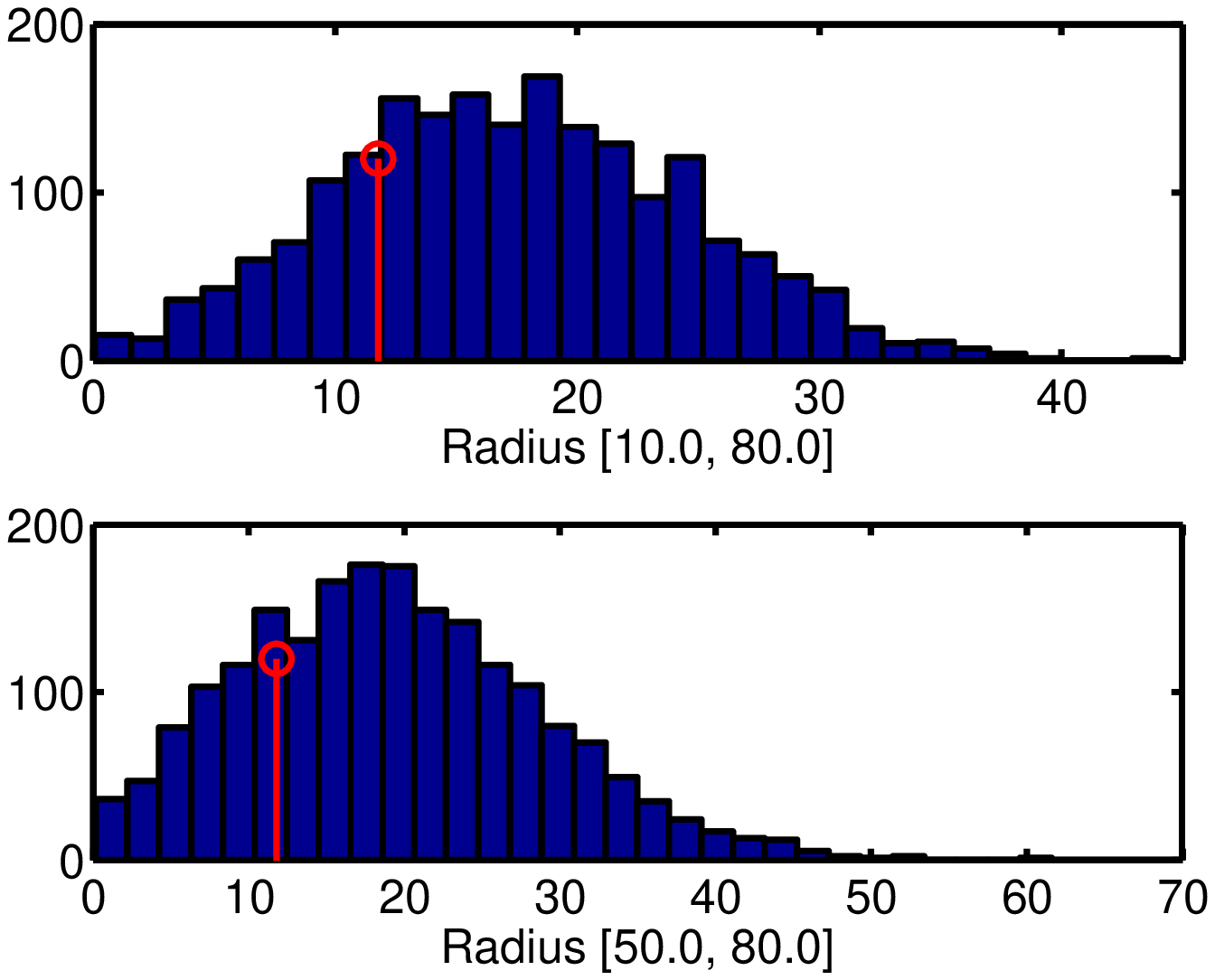}
	}
	\caption{Comparison of posterior distributions under different settings of the prior parameters (run on synthetic data). The top row results from the prior $\mathcal{N}(10, 80)$, and the bottom row uses $\mathcal{N}(50, 80)$.}
\label{fig:priors}
\end{figure}

Next, we investigate the effect of the prior parameters. As we mentioned, our priors are set to be rather uninformative. That is, we set a large variance to encode our uncertainty of the values of these parameters. We generated $10$ synthetic networks using \textit{Radius+Comms} model's generative process (after distributing nodes uniformly over a given region of space) so that we know the true parameter values. Then, we ran our inference algorithm on the observed networks using different settings for the prior distributions. Figure~\ref{fig:priors} shows the resulting posterior distributions, as well as the generating parameter values, for one synthetic network.

For all parameters, the top and bottom rows show the posterior distribution when the prior mean was set to $10$ and $50$ respectively. The prior variance was kept at $80$ to capture our prior uncertainty in these parameters. For both settings of the prior, we see that all of the posteriors are centered around the the parameter value with which the observed networks were generated. We do notice a rather slight shift in the posterior when the prior mean was set to $50$, though the mode still converges to the correct area. From this analysis, we show that the priors have little effect on the posterior, though they do play a role in convergence.

\section{Analysis of the \textit{C. elegans} Network}
\label{sec:discussion}
In the previous section, we showed that our proposed models provide an accurate fit to several real world spatial networks. Next, we analyze the inferred parameter values for \textit{Radius+Comms} on the \textit{C. elegans} network. We focus on \textit{C. elegans} because detailed information about the nodes (i.e. neurons) is available, thus we are best able to interpret and explain our findings~\cite{Wormatlas}.

\begin{figure}[ht]
	\centering
	\includegraphics[keepaspectratio, width=0.45\textwidth]{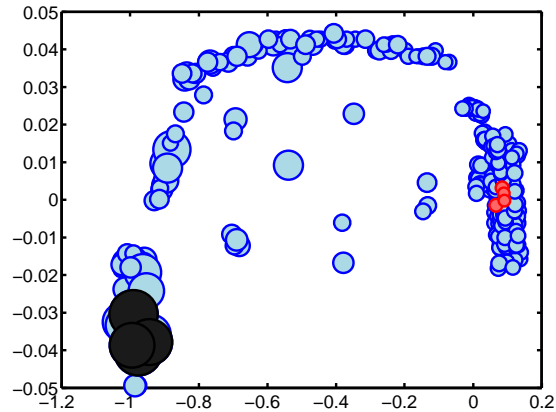}
	\caption{Analysis of radii from the \textit{C. elegans} network.}
	\label{fig:celegans_net}
\end{figure}

We first analyze the relationship between radius and a node's position within the network. Figure~\ref{fig:celegans_net} shows the location and mean posterior radius for each node in the \textit{C. elegans} network. Note radii are scaled for easier visualization, thus the node size captures relative differences in the size of the radius, not the absolute magnitude. We highlight the nodes with the largest (top-$4$ are shown in black) and smallest (shown in red) radii.

\begin{figure}[ht]
    \centering
    \subfigure[\textit{Radius}]{
	\label{fig:base_radii}
	\centering
	\includegraphics[keepaspectratio, width=0.22\textwidth]{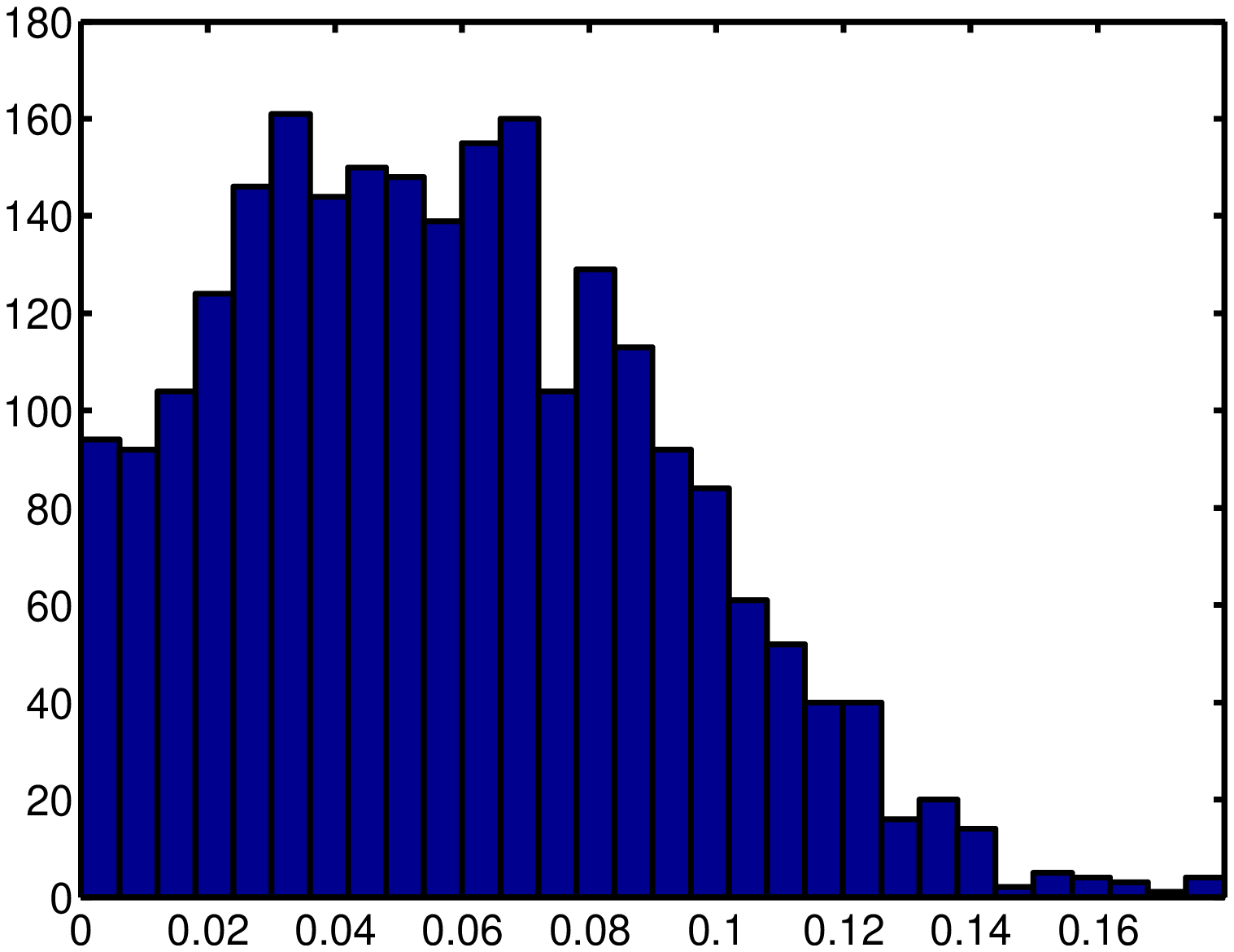}
    }
    \subfigure[\textit{Radius+Comms}]{
	\label{fig:context_radii}
	\centering
	\includegraphics[keepaspectratio, width=0.22\textwidth]{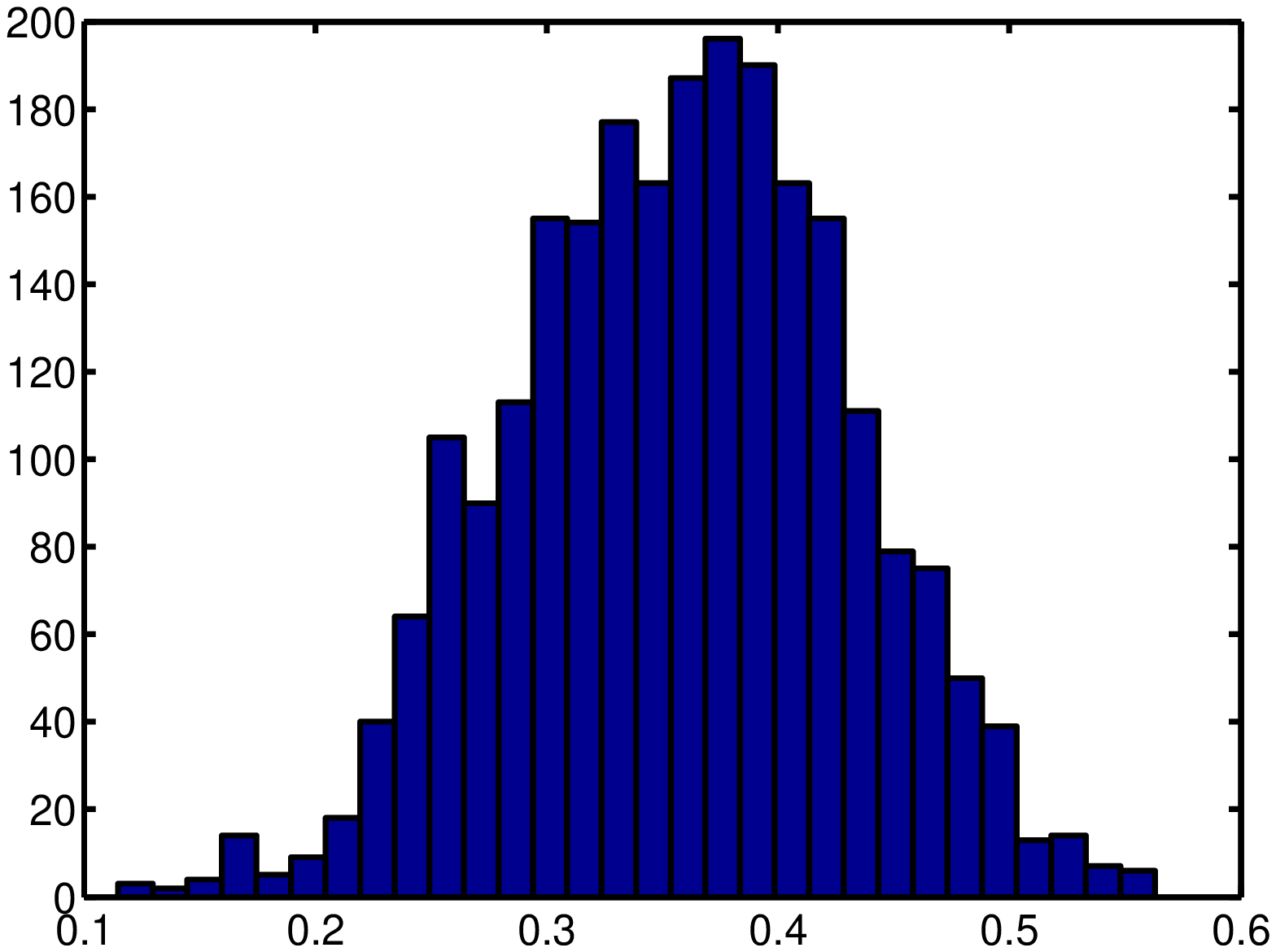}
    }
    \caption{Posterior samples of the radius for the neuron PVCL, which has one of the largest (posterior average) radii in the network (in both models).}
    \label{fig:celegans_postr}
\end{figure}

The neurons with the largest radii are PVC[L/R] and DV[A/B]. The DVA neuron functions in mechanosensory integration, providing input to both the anterior and posterior touch circuits~\cite{Wormatlas}. Neurons taking part in such sensory integration naturally need to interact with a wide variety of spatially disperse neurons in order to collect this information, thus explaining the need for a large spatial reach. The PVC[L/R] neurons are known to form synapses with the VB group of neurons (motor neurons) which are located in the head of the worm, as well as the DB neurons (dorsal motor neurons) which are located throughout the body of the worm. Given that the PVC[L/R] neurons are located in the tail, they must extend a long distance to form these links. We show histograms of the posterior distribution of the radius of PVCL for each of the models in figure~\ref{fig:celegans_postr}.

\begin{figure}[ht]
    \centering
    \includegraphics[keepaspectratio, width=0.45\textwidth]{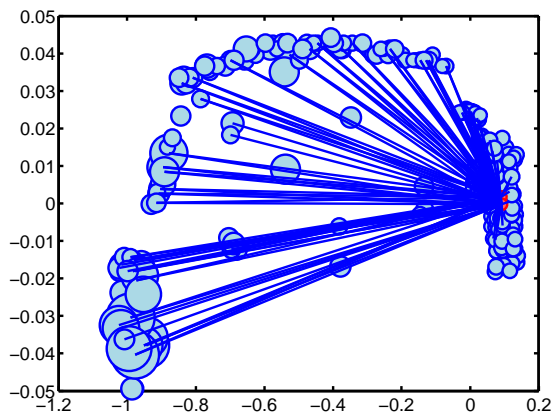}
    \caption{Connections formed by the AVA neurons (shown in red).}
    \label{fig:celegans_ava}
\end{figure}

The smallest radii belong to the AVE[L/R] and AVA[L/R] neurons, all of which are located in the head of the worm. Interestingly, it is known that the processes (axons and dendrites) of the AVE[L/R] neurons are restricted to the area above the vulva, which is typically found near the center of the worm body~\cite{Donald1997, Wormatlas}. This limited spatial reach, combined with the fact that the neurons lie in the head of the worm, where neurons are most dense, explain this node's small radius. In contrast, the AVA[L/R] neurons are the pair with the largest degrees, with 76 and $74$ connections respectively. Moreover, these neurons run the entire length of the ventral nerve cord as they function in forward and backward movement~\cite{Donald1997, Wormatlas}. Given the wide reach of these neurons, it seems peculiar that they would not have larger radii. However, upon further inspection, we see that although they form many connections with neurons spread throughout the body of the worm, they also neglect to form connections with many neurons in the head (see figure~\ref{fig:celegans_ava}). Because there is a high density of neurons in the head of the worm, if these neurons do not form connections with other neurons in this region, their radii will be penalized heavily. Thus, many neurons in this area have very small \textit{spatial reach} and other nodes in less dense regions are forced to increase their \textit{spatial reach} to pick up the slack.

\section{Conclusions}
\label{sec:conclusions}
We have introduced a novel approach for modeling spatial networks which utilizes a node-centric spatial cost function. To learn this function, we attach a latent radius parameter to each node, which describes the spatial reach of that node, thus summarizing the local network structure surrounding that node. Additionally, we have provided a natural extension to this model which captures salient community structure, which cannot be explained due to spatial or node popularity effects.

We have shown experimentally that our models, \textit{Radius} and \textit{Radius+Comms}, result in higher quality link predictions across the different datasets than competing techniques. Interestingly, the most substantial improvements came from predicting links between nodes with low observed degrees. That is, the nodes from which the network structure provides the least amount of information. Furthermore, we analyze the model parameters and offer interpretations of the inferred values on the test networks.

Studying the role of space in networks is critical to further our understanding of complex systems. In this work, we have introduced a model which offers the flexibility required to appropriately account for complicated link-distance cost functions as well as other connection properties. Our model provides a node-centric view of the unobserved link-distance cost function which influences the network structure. This approach offers greater modeling flexibility, and, as we have demonstrated, a more accurate representation of the data.

{\small
  \bibliographystyle{plos2009}
}

\end{document}